\documentclass[10pt,final,3p,times]{elsarticle}
\usepackage{amsthm} 
\usepackage{lipsum}
\usepackage{amsfonts}
\usepackage{graphicx}
\usepackage{natbib}
\usepackage{hyperref}
\usepackage{epsfig}
\usepackage{amssymb}
\usepackage{cite}
\usepackage{float}
\usepackage{bigints}
\usepackage{amssymb}
\usepackage{mathtools}
\newtheorem{definition}{Definition}[section]
\begin{document}
\begin{frontmatter}
\title{A Novel Approach to find Exact Solutions of Nonlinear ODE Systems: Applications to Coupled Non-Integrable and Integrable mKdV Equations  
}
\author{Prakash Kumar Das}
\ead{prakashdas.das1@gmail.com}


\address{Department of Mathematics, Trivenidevi Bhalotia College, Raniganj 713347, West Bengal, India}

\begin{abstract}
A novel approach is introduced for deriving exact solutions to nonlinear systems of ordinary differential equations. This method consists of four parts. In the initial part, the examined nonlinear differential equation system is transformed into a linear differential equation system using an infinite series sum transformation and Adomian polynomials. In the second part, we presented a modified method of variation of parameters utilizing a revised Cramer's rule for addressing a linear system of block matrices. In the following part, we merged the outcomes of the prior two parts and introduced a new recursive method for acquiring a series solution of the system established in the first section. In the final part, we introduce the fundamental ideas of the multiplicative inverse of power series and methods for summing infinite series. The method utilizes the property that the multiplicative inverse of the power series either terminates after a finite number of terms or simplifies to a geometric series, which can be easily summed to yield closed-form solutions for the series in question. The effectiveness of this method is shown by successfully deriving exact localized solutions for both integrable and non-integrable coupled modified Korteweg-de Vries equations. Using the multiplicative inverse technique, it is shown that a rational generating function can be created for the series solution derived from the proposed method for the specified system of nonlinear differential equations, regardless of whether it is integrable or non-integrable. \\
\it{ \textbf{Keywords:} Variation of parameters method, boundary value problems, Adomian polynomials, nonlinear equations, exact solutions, multiplicative inverse of the power series,  mKdV Equations.}
\end{abstract}
\end{frontmatter}
\section{Introduction}
The method of variation of parameters (MVP) was independently developed by Leonhard Euler in 1748 and Joseph Louis Lagrange in 1774. It is a technique used to transform solutions of linear homogeneous ordinary differential equations into particular solutions for related inhomogeneous equations. This method enhances the classical order reduction technique for solving nonhomogeneous linear differential equations. Recently, MVP has been successfully applied beyond linear cases, effectively deriving approximate numerical solutions for nonlinear boundary value problems, including micropolar flow \citep{gungor2020application} and complex heat conduction-radiation problems \citep{moore2014application}. Various studies show MVP yielding accurate numerical solutions for nonlinear sixth-order boundary value problems  \citep{noor2008variation,mohyud2009variation}. Compared to semi-analytical methods like the Adomian decomposition method and differential transformation method, MVP has proven more effective and less sensitive to the degree of nonlinearity, making it a strong alternative for engineering boundary value problems where efficient solutions are critical \citep{moore2019comparison}. Thus, MVP is a versatile and efficient method for solving both linear and nonlinear differential equations in a variety of scientific and engineering contexts.
In all the previous studies, the classic method of variation of parameters (MVP) used definite integrals  hence they only obtained approximate analytical/numerical solutions.
From the preceding discussion, it is evident that the  MVP with definite integration has thus far been utilized on nonlinear equations to derive numerical and approximate solutions. Motivated by its precision and achievements in surpassing other techniques, we are compelled to inquire if this method can be adjusted and utilized to obtain exact closed-form solutions of nonlinear scalar and system of differential equations as well? To answer the question for scalar case, in contrast to this traditional view, in a recent study the author proposed  a modified method of variation of parameters (MMVP)  \citep{das2025new} involving indefinite integrals that effectively produces exact closed-form solutions for nonlinear ordinary differential equations. This new approach yielded solutions expressed in terms of exponential, hyperbolic, trigonometric, algebraic, and Jacobi elliptic functions, as demonstrated in applications such as thermophoretic wave modeling in graphene sheets. 
This study seeks to deliver the answer to that inquiry for system of equations. Inspired by these advancements, the current work proposes a modification of  MMVP and applied  for the first time to systems of nonlinear differential equations in search of exact closed-form solutions. Thus, while the classical MVP relies heavily on numerical evaluation of definite integrals leading to approximations in nonlinear cases, the MMVP offers a promising pathway to derive exact, closed-form solutions for nonlinear differential equations.

The main concern of any series solution method that deals with exact solutions is that it is not always possible to find a closed form or generating function for the derived series. It is not always straightforward to find the closed form of the series solution of integrable equations. However, it becomes impossible to find the closed form of the series solution of non-integrable equations. Therefore, it is a general question to ask whether there exists any technique that easily provides the closed form of the derived series solutions for integrable equations and, under certain restrictions, for non-integrable equations. Inspired by a recent study on power series \citep{Zemlyanukhin}, we believe that the multiplicative inverses of power series could answer the above question after combining a few techniques with them. Therefore, in this study, we use the multiplicative inverses of power series along with a few techniques for summing series. We aim to show that this combination can provide the closed-form solutions of derived power series for integrable and non-integrable equations.

The detailed steps for employing MMVP to solve a system of higher order nonlinear differential equations are covered in the section \ref{MVP}. The process of reducing a finite system of higher order nonlinear differential equations to a system of infinite linear equations is illustrated in the subsection \ref{red}. The particular procedures of MMVP for solving the linear system of differential equations shown in earlier section are covered in subsection \ref{sMVP}. Section \ref{MVP}'s last subsection, \ref{sSystem}, shows how to combine the results of previous sections to arrive at a definitive recursive formula for solving the given higher order nonlinear differential equation system. Section \ref{mi} discusses the fundamental ideas of the  multiplicative inverse of a power series as well as techniques for determining  generating function of a a power series.     The usage of MMVP to solve the system of integrable and non-integrable mKdV equations is covered in Section \ref{Appl}. Additionally covered here is how to use the multiplicative inverse of power series approach to find closed-form solutions for the derived power series. Here, using MMVP, we have effectively obtained closed-form solutions for the given equation containing exponential functions. The section \ref{com} discusses MMVP's comparison with other approaches.  The advantages and disadvantages of the suggested approach are covered in Section \ref{AdLim}.  Section \ref{con} concludes with a discussion of our findings.

\section{MMVP for higher order nonlinear system of differential equations}\label{MVP}
\subsection{Reduction of higher order nonlinear system of differential equations to system of linear equations}\label{red}
Consider the subsequent $n$-th order nonlinear system comprising $m$ differential equations:
\begin {eqnarray}\label{eq1p1}
{\cal P}_0(x).\ {\bf y }^{(n)}(x) + {\cal P}_1(x).\ {\bf y}^{(n-1)}(x) +\cdots+ {\cal P}_n(x).\ {\bf y}(x) = {\cal N}[{\bf y}](x), \, \, \, \, x \in \mathbb{R}.
\end{eqnarray}
 Here ${\bf y }$ is $m\times 1$  matrix of dependent variables, ${\cal P}_0(x), {\cal P}_1(x)\ \cdots, {\cal P}_n(x)$ are continuous coefficient matrices  of order $m\times m$, and  ${\cal N}[y](x)$ is a matrix of nonlinear terms of order $m \times 1$.  
We will abbreviate Equation Eq. (\ref{eq1p1}) as 
\begin {eqnarray}\label{eq1p2a}
\hat{{\cal L}}[{\bf y }](x)=  {\cal N}[{\bf y }](x),
\end{eqnarray}
where
\begin {eqnarray}\label{eq1p2}
\hat{{\cal L}}[{\bf y }](x)= {\cal P}_0(x).\ {\bf y }^{(n)}(x) + {\cal P}_1(x).\ {\bf y}^{(n-1)}(x) +\cdots+ {\cal P}_n(x).\ {\bf y}(x). 
\end{eqnarray}
Eq. (\ref{eq1p2a}) can be transformed into a linear system equation by writing
\begin{equation}\label{eq1p3}
{\bf y }(x) = \sum_{k=0}^\infty \epsilon^{\frac{k}{2}} \, {\bf y }_{k}(x)
\end{equation}  
 and  rewrite  the nonlinear term into the series 
\begin{equation}\label{eq1p4}
{\cal N}[{\bf y }](x) = \epsilon^{p} \, \sum_{k=0}^\infty \epsilon^{\frac{k}{2}} {\cal A}_k({\bf y }_{0}(x),{\bf y }_{1}(x),\cdots ,{\bf y }_{k}(x))\equiv \epsilon^{p} \, \sum_{k=0}^\infty \epsilon^{\frac{k}{2}} \, {\cal A}_k(x),
\end{equation}
where ${\bf y }_{k}(x)$ are matrices of order $m \times 1$, and  ${\cal A}_k(x) \ (\ k\geq 0) $ are $m \times 1$ matrix of  Adomian polynomials \citep{adomian2013solving,das2018solutions}. These polynomials can be systematically derive by applying the formula
\begin {equation}\label{eq1p5}
{\cal A}_k(x)= \frac{1}{k!}\left[\frac{d^k}{d \bar{\epsilon}^k}{\cal N} \left( \sum_{l=0}^\infty {\bf y }_l(x) \ \bar{\epsilon}^l \right)\right]_{\bar{\epsilon}=0}, \ \ k \geq 0.
\end{equation}
Note that the $\epsilon$ and $\bar{\epsilon}$ symbols in (\ref{eq1p3})-(\ref{eq1p5}) do not correspond to any physical perturbation parameter. Instead, they act as stand-ins for nonlinear terms in later iterations.
By applying (\ref{eq1p3}) and (\ref{eq1p4}) to (\ref{eq1p2a}), we get
\begin {eqnarray}\label{eq1p6}
 \sum_{k=0}^\infty \epsilon^{\frac{k}{2}} \, \hat{\cal L}\left[{\bf y }_{k}(x) \right] =\epsilon^{p} \, \sum_{k=0}^\infty \epsilon^{\frac{k}{2}} \, {\cal A}_k(x).
\end{eqnarray}
Interestingly, replacing $p = \frac{1}{2}$ in (\ref{eq1p6}) with the coefficients of $\epsilon$ yields a system of linear equations 
\begin{equation}\label{eq1p7}
  \hat{\cal L}\left[{\bf y }_{0}(x) \right]=0, \ \ 
  \hat{\cal L}\left[{\bf y }_{k}(x) \right] = {\cal A}_{k-1}(x),\ \ k \geq 1
\end{equation}
including non-homogeneous terms ${\cal A}_{k-1}(x)$ and matrices of dependent variables ${\bf y }_{k}(\xi)$.
On the other hand, choosing $p=1$ in (\ref{eq1p6}) yields the linear system
\begin{equation}\label{eq1p8}
  \hat{\cal L}\left[{\bf y }_{0}(x) \right]=0, \ \
  \hat{\cal L}\left[{\bf y }_{1}(x) \right]=0, \ \
   \hat{\cal L}\left[{\bf y }_{k}(x) \right] = {\cal A}_{k-2}(x),\ \ k \geq 2.
\end{equation}
To solve the system of equations (\ref{eq1p7}) and (\ref{eq1p8}), we will propose a modified method of variation of parameters in the next section.
\subsection{Solution of system of linear differential equations (\ref{eq1p7}) and (\ref{eq1p8}) by MMVP}\label{sMVP}
This section deals with the nonhomogeneous system of equation
\begin {eqnarray}\label{eq2p1}
\hat{\cal L}\left[{\bf y }_{k}(x) \right] ={\cal F}_k(x).
\end{eqnarray}
and  use ${\cal F}_k(x)={\cal A}_{k-1}(x)$ when addressing system (\ref{eq1p7}) and ${\cal F}_k(x)={\cal A}_{k-2}(x)$ for system (\ref{eq1p8})For system (\ref{eq1p7}), we use ${\cal F}_k(x)={\cal A}_{k-1}(x)$, and for system (\ref{eq1p8}), we use ${\cal F}_k(x)={\cal A}_{k-2}(x)$. 
Importantly, all solutions to this equation and its complementary equation $\hat{\cal L}\left[{\bf y }_{k}(x) \right] = 0$ are considered in $ \mathbb{R}$.  Presently, the complementary solution of the linear portion of every equation is $\hat{\cal L}\left[{\bf y }_{k}(x) \right] = 0\ i=0, 1, 2, \cdots $  has the following form
\begin {eqnarray}\label{eq2p2}
{\bf y }_{k}^{c}(x) =   \tilde{{\bf y} }_{1}(x).\ {\bf u}_1  +   \tilde{{\bf y}}_{2}(x).\ {\bf u} _2  +\cdots+  \tilde{{\bf y}}_{n}(x). \ {\bf u}_n ,
\end{eqnarray}
where ${\bf u}_i,\ i=1,2,\cdots n$ are arbitrary constant matrices of order $m \times 1$. The set of solutions $\{ \tilde{{\bf y} }_{1}(x), \tilde{{\bf y} }_{2}(x), \cdots, \tilde{{\bf y} }_{n}(x)\}$ are referred to as fundamental matrix solutions of $m \times m$ ordered matrices.  
To find a specific solution of (\ref{eq2p1}), we will show how to use the MMVP when we know a basic set of solutions $\{ \tilde{{\bf y} }_{1}(x), \tilde{{\bf y} }_{2}(x), \cdots, \tilde{{\bf y} }_{n}(x)\}$.
Our attempt aims to find a specific solution to (\ref{eq2p1}) in the form of 
\begin {eqnarray}\label{eq2p3}
{\bf y }_{k}^{p}(x) =   \tilde{{\bf y} }_{1}(x).\ {\bf u}_1(x)  +  \tilde{{\bf y}}_{2}(x). \ {\bf u} _2(x)  +\cdots+  \tilde{{\bf y}}_{n}(x). \ {\bf u}_n(x) 
\end{eqnarray}
where the matrices of order $m \times 1$ that have not yet been recognized are $\{ {\bf u}_1(x), {\bf u}_2(x), \cdots, {\bf u}_n(x)\}$. The following $(n-1)$ restrictions are first applied to $\{ {\bf u}_1(x), {\bf u}_2(x), \cdots, {\bf u}_n(x)\}$: 
\begin {eqnarray}\label{eq2p5}
\begin{cases}
  \tilde{{\bf y} }_{1}(x). \ {\bf u}_1'(x)  +    \tilde{{\bf y} }_{2}(x).\ {\bf u}_2'(x)  +\cdots+  \tilde{{\bf y} }_{n}(x).\ {\bf u}_n'(x) = 0,\\
  \tilde{{\bf y} }_{1}'(x). {\bf u}_1'(x)  +  \tilde{{\bf y} }_{2}'(x). \ {\bf u}_2'(x)  +\cdots+  \tilde{{\bf y} }_{n}'(x).\ {\bf u}_n'(x) = 0, \\
      \hspace{1in} \vdots  \\
  \tilde{{\bf y} }_{1}^{(n-2)}(x). \ {\bf u}_1'(x)  +   \tilde{{\bf y} }_{2}^{(n-2)}(x).\ {\bf u}_2'(x)  +\cdots+   \tilde{{\bf y} }_{n}^{(n-2)}(x).\ {\bf u}_n'(x) = 0. 
\end{cases}
\end{eqnarray}
Under these conditions, the higher derivatives of ${\bf y}_k(x)$ have simple equations:
\begin {eqnarray}\label{eq2p6}
{\bf y}_k^{(r)}(x) =    \tilde{{\bf y} }_{1}^{(r)}(x).\ {\bf u}_1(x)  +    \tilde{{\bf y} }_{2}^{(r)}(x).\ {\bf u}_2(x) +\cdots+  \tilde{{\bf y} }_{n}^{(r)}(x).\ {\bf u}_n(x), \  \ 0 \leq r \leq (n-1).
\end{eqnarray}
The last equation in equation (\ref{eq2p6}) is 
\begin {eqnarray}\label{eq2p6a}
{\bf y}_k^{(n-1)}(x) =   \tilde{{\bf y} }_{1}^{(n-1)}(x).\ {\bf u}_1(x)  +  \tilde{{\bf y} }_{2}^{(n-1)}(x).\ {\bf u}_2(x) +\cdots+    \tilde{{\bf y} }_{n}^{(n-1)}(x).\ {\bf u}_n(x).
\end{eqnarray}
Calculating the derivative of this yields 
\begin {eqnarray}\label{eq2p7}
y_k^{(n)}(x) =  &&  \tilde{{\bf y} }_{1}^{(n)}(x).\ {\bf u}_1(x)  +  \tilde{{\bf y} }_{2}^{(n)}(x).\ {\bf u}_2(x) +\cdots+    \tilde{{\bf y} }_{n}^{(n)}(x).\ {\bf u}_n(x) \nonumber \\ && +
  \tilde{{\bf y} }_{1}^{(n-1)}(x).\ {\bf u}_1'(x) +    \tilde{{\bf y} }_{2}^{(n-1)}(x).\ {\bf u}_2'(x) +\cdots+    \tilde{{\bf y} }_{n}^{(n-1)}(x).\ {\bf u}_n'(x).
\end{eqnarray}
This and equation (\ref{eq2p6}) are inserted into equation (\ref{eq2p1}) to get
\begin {eqnarray}\label{eq2p8}
 &&  \hat{\cal L}\left[  \tilde{{\bf y} }_{1}\right](x).\ {\bf u}_1(x)  +  \hat{\cal L}\left[  \tilde{{\bf y} }_{2}\right](x).\ {\bf u}_2(x) +\cdots+ \hat{\cal L}\left[  \tilde{{\bf y} }_{n}\right](x).\ {\bf u}_n(x) \nonumber \\
 && +  {\cal P}_0(x).\ \left(    \tilde{{\bf y} }_{1}^{(n-1)}(x).\ {\bf u}_1'(x)  +    \tilde{{\bf y} }_{2}^{(n-1)}(x).\ {\bf u}_2'(x) +\cdots+    \tilde{{\bf y} }_{n}^{(n-1)}(x).\ {\bf u}_n'(x) \right) = {\cal F}_k(x).
\end{eqnarray}
Considering $\hat{\cal L}\left[  \tilde{{\bf y} }_{i}\right](x) = 0, \ (1\leq i \leq n)$, we get
\begin {eqnarray}\label{eq2p9}
  \left(    \tilde{{\bf y} }_{1}^{(n-1)}(x).\ {\bf u}_1'(x)  +    \tilde{{\bf y} }_{2}^{(n-1)}(x).\ {\bf u}_2'(x) +\cdots+    \tilde{{\bf y} }_{n}^{(n-1)}(x).\ {\bf u}_n'(x) \right) ={\cal P}_0(x)^{-1}.\ {\cal F}_k(x).
\end{eqnarray}
 It shows that if conditions (\ref{eq2p5}) and (\ref{eq2p9}) are satisfied, then (\ref{eq2p3}) is a solution to equation (\ref{eq2p1}). It should be noted that the combination of Equations (\ref{eq2p5}) and (\ref{eq2p9}) can be represented in block matrix form as
\begin {eqnarray}\label{eq2p10}
 \left[\begin{array}{cccc}
   \tilde{{\bf y} }_{1}(x)&   \tilde{{\bf y} }_{2}(x)&\cdots &   \tilde{{\bf y} }_{n}(x) \\   \tilde{{\bf y} }_{1}'(x)&   \tilde{{\bf y} }_{2}'(x)&\cdots &   \tilde{{\bf y} }_{n}'(x)   \\ \vdots & \vdots &\ddots & \vdots \\   \tilde{{\bf y} }_{1}^{(n-2)}(x)&   \tilde{{\bf y} }_{2}^{(n-2)}(x)&\cdots &   \tilde{{\bf y} }_{n}^{(n-2)}(x)  \\    \tilde{{\bf y} }_{1}^{(n-1)}(x)&   \tilde{{\bf y} }_{2}^{(n-1)}(x)&\cdots &   \tilde{{\bf y} }_{n}^{(n-1)}(x)  \\\end{array}\right]\left[\begin{array}{c} {\bf u}_1'(x)\\ {\bf u}_2'(x)\\ \vdots \\ {\bf u}_{n-1}'(x)  \\ {\bf u}_n'(x) \\ \end{array}\right]=\left[\begin{array}{c}
 0\\ 0\\ \vdots \\ 0  \\ {\cal P}_0(x)^{-1}.\ {\cal F}_k(x) \\ \end{array}\right].
 \end{eqnarray}
where the first matrix is a block Wronskian matrix of order $(m n)\times (m n)$, while the second and third are block matrices of order $(m n) \times 1$.
To solve the system of equations of block matrices, a minor adjustment to Cramer's rule \citep{brunetti2014old,huang2025new} yields 
\begin {eqnarray}\label{eq2p11}
{\bf u}_{l}'(x) =  \frac{ 1}{ W(x)} \ \Bigg( (-1)^{i+ j} W_{i,j}(x)\Bigg)^T _{\substack{i = m n - (m-1),\ \cdots ,\ m n  \\ j = (l-1)m + 1,\ \cdots ,\ l m }}.\left({\cal P}_0(x)^{-1}.\ {\cal F}_k(x)\right), \ \ \ \ 1 \leq l \leq n,
\end{eqnarray}
where $W(x)$ is the Wronskian of the fundamental solution set $\{  \tilde{{\bf y} }_{1}(x),  \tilde{{\bf y} }_{2}(x), \cdots,  \tilde{{\bf y} }_{n}(x) \}$, which is non-zero on $ \mathbb{R}$.  In addition, $\Bigg( (-1)^{i+ j} W_{i,j}(x)\Bigg)^T$ is the transpose of a $m \times m$ matrix with signed elements $W_{i,j}(x)$, which are the determinant obtained by removing the $i-$th row and the $j-$th column from $W(x)$ for $i = m n - (m-1),\ \cdots,\ m n$, and $ j = (l-1)m + 1,\ \cdots,\ l m$.
After obtaining $ {\bf u}_1'(x), {\bf u}_2'(x), \cdots, {\bf u}_n'(x)$, we can use integration to obtain $ {\bf u}_1(x), {\bf u}_2(x), \cdots, {\bf u}_n(x)$. We take the integration constants to be zero.  This  delivers  the solution to equation (\ref{eq2p1}) in the format 
\begin {eqnarray}\label{eq2p12}
{\bf y}_k(x) = \sum_{l=1}^{n} \left\{   \tilde{{\bf y} }_{l}(x). \bigintsss \left[ \frac{ 1}{ W(x)} \ \Bigg( (-1)^{i+ j} W_{i,j}(x)\Bigg)^T _{\substack{i = m n - (m-1),\ \cdots ,\ m n  \\ j = (l-1)m + 1,\ \cdots ,\ l m }}.\left({\cal P}_0(x)^{-1}.\ {\cal F}_k(x)\right) \right] dx \right\}.   
\end{eqnarray}
\subsection{Solution of system of equations (\ref{eq1p7}) and (\ref{eq1p8})}\label{sSystem}
In order to use MMVP to address the system (\ref{eq1p7}), we must solve the recursive system
\begin {eqnarray}\label{eq2p13}
\begin{cases}
{\bf y}_0(x) = \sum_{l=1}^{n}  \tilde{{\bf y} }_{l}(x).c_{0,l} ,  \\
{\bf y}_k(x) = \sum_{l=1}^{n} \left\{   \tilde{{\bf y} }_{l}(x). \bigintsss \left[ \frac{ 1}{ W(x)} \ \Bigg( (-1)^{i+ j} W_{i,j}(x)\Bigg)^T _{\substack{i = m n - (m-1),\ \cdots ,\ m n  \\ j = (l-1)m + 1,\ \cdots ,\ l m }}.\left({\cal P}_0(x)^{-1}.\ {\cal A}_{k-1}(x)\right) \right] dx \right\},   \ \ k \geq 1,
\end{cases}
\end{eqnarray}
where $\tilde{{\bf y} }_{j}(x)$ are the solutions to $\hat{\cal L}\left[{\bf y}_{k}(x) \right] = 0.$ and $c_{0,j}, \ j=1,2, \cdots n,$ are selected constants. This leads to the series solution of equation (\ref{eq1p1}) represented as follows
\begin{equation}\label{eq2p14}
{\bf y}(x) = \sum_{k=0}^\infty  {\bf y}_{k}(x).
\end{equation} 
Similarly, system (\ref{eq1p8}) can be used to obtain a series solution in the form of (\ref{eq2p14}) by employing the recursive system
\begin {eqnarray}\label{eq2p15}
\begin{cases}
 {\bf y}_0(x) = \sum_{l=1}^{n}   \tilde{{\bf y} }_{l}(x).c_{0,l}, \\
{\bf y}_1(x) = \sum_{l=1}^{n}  \tilde{{\bf y} }_{l}(x).c_{1,l}, \\ 
{\bf y}_k(x) = \sum_{l=1}^{n} \left\{   \tilde{{\bf y} }_{l}(x). \bigintsss \left[ \frac{ 1}{ W(x)} \ \Bigg( (-1)^{i+ j} W_{i,j}(x)\Bigg)^T _{\substack{i = m n - (m-1),\ \cdots ,\ m n  \\ j = (l-1)m + 1,\ \cdots ,\ l m }}.\left({\cal P}_0(x)^{-1}.\ {\cal A}_{k-2}(x)\right) \right] dx \right\}, \ \ k \geq 2, 
\end{cases}
\end{eqnarray}
where $c_{l,j}, \ l=0,1$ are arbitrary constants and $ \tilde{{\bf y} }_{j}(x)$ are solutions to $\hat{\cal L}\left[{\bf y}_{k}(x)\right] = 0$ for  $j=0,1,2,\cdots,n.$

It is essential to point out that we have set the integration constants to zero during the integration of \( {\bf u}_j'(x), \ j=1, 2, \cdots, n \), in (\ref{eq2p12}). If we retain them, they will be absorbed into the terms of \( {\bf y}_0(x) \) of the series solution (\ref{eq2p14}) without yielding any significantly new results. It is essential to recognize that utilizing symbolic computations often enables us to accurately determine the general term or generating functions of the iterative scheme (\ref{eq2p13}) or (\ref{eq2p15}). Whenever we can determine the general term or generating functions, the series sum (\ref{eq2p14}) consistently provides a closed-form solution to the nonlinear differential equation (\ref{eq1p1}). However, we use the multiplicative inverse of power series method when we are unable to identify the general term or generating functions of the series sum (\ref{eq2p14}). The nonlinear differential equation (\ref{eq1p1}) can be solved in closed form if the multiplicative inverse power series ends after finite terms or if it gives the closed form of the series (directly or with specific parameter restrictions). We impose limits on the series terms or use parameters to truncate the series in order to discover the closed form solution if it does not converge or truncate. The fundamental ideas of the multiplicative inverse of power series and the truncation or summing technique of these series are covered in the next section. 
\section{Multiplicative inverse of power series}\label{mi}
\begin{definition}
\textbf{(Multiplicative inverse of power series)} A formal power series $ A(\zeta) = \sum_{n=0}^\infty a_n \ \zeta^n $ in a commutative ring $ R $ (such as $ \mathbb{R} $ or $ \mathbb{C} $), said to have a multiplicative inverse in the form $ (A(\zeta))^{-1} = \sum_{n=0}^\infty b_n \  \zeta^n $
if  $ A(\zeta)\cdot (A(\zeta))^{-1} = 1 $ and $ a_0 $ is invertible (that is, $ a_0 \neq 0 $ ) in $ R $ \citep{Zemlyanukhin}. 
\end{definition}
One can use the following recursive formula to determine the coefficients of a multiplicative inverse power series:
\begin{eqnarray}\label{mi1}
&& b_0 = \frac{1}{a_0},  \nonumber \\
&& b_n = -\frac{1}{a_0} \sum_{k=1}^{n} a_k \ b_{n-k} \ \  \text{ for all } \ \   n \geq 1.
\end{eqnarray}
In this article, we determine the multiplicative inverse of power series in cases where the original series cannot be summated. Therefore, summability of multiplicative inverse is crucial and required in order to determine the close form of the original series. When the equation under consideration is integrable, the multiplicative inverse of the acquired series solution reduces to a straightforward infinite series, the sum of which can be manually calculated or truncated after certain terms. However, the derivation of the closed form of the deduced series solution using MMVP becomes complicated whenever the considered equation is not integrable. By imposing restrictions on the terms of the series, we can either compel the multiplicative inverse of the produced series to truncate after finite terms or make it comparable with a convergent infinite geometric series whose closed form is known. These constraints frequently resulted in integrable conditions on the parameters of the equations under consideration or made the derived series convergent. These conditions therefore lead us to specific exact solutions of the non-integrable equations under consideration. The next section discusses the methods used to impose specific limitations on the series. 
\subsection{Summing or truncating technique of a multiplicative inverse of power series}
Most of the time, determining the series' close form is not simple. Therefore, we use each of these techniques to determine the closed form or generating function of the multiplicative inverse of a power series or original power series.\\
1) \textbf{ Truncating the series:}  We set the coefficients ($b_n$) after a specific term to zero in order to truncate infinite series after finite terms. Series end after a finite term if all of the after terms share a common factor. This is because equating this common factor with zero imposes some restrictions on the parameters involved.  If these factors cannot be found directly, we set the coefficients of the terms that follow to zero and see if the remaining terms in this series vanish or not. By doing this, we are able to determine an appropriate restriction on the parameters for which the series ends after finite terms.\\
  2) \textbf{ Summing infinite series:}  These infinite series are forced to become convergent geometric series, or series with a common ration, in order to extract the generating function.  We suppose that the corresponding series have a common ration for this, so that $$\frac{b_{n+1}}{b_n}=\frac{b_{n+2}}{b_{n+1}}, \ n \geq 1. $$ Solving the above criteria frequently results in a convergent geometric series with a common ratio, which places some restrictions on the parameters involved.
\section{Application of MMVP}\label{Appl}
In order to describe the procedures and effectiveness of our suggested approach, we examine two systems of mKdV equations in this section: one that is integrable and one that is not, and we solve both using MMVP. 
\subsection{Example 1:} Consider the following generalization of modified KdV (mKdV) equations, extended to multiple interacting fields \citep{wazwaz2012study,mahmood2014tan}
\begin{eqnarray}\label{ex2eq1}
&& u_t(x,t) + u_{xxx}(x,t) + \left(6 u^2(x,t) u_x(x,t) + 3 u(x,t) \ v(x,t)\  w_x(x,t) \right) = 0, \nonumber
 \\
&& v_t(x,t) + v_{xxx}(x,t) + \left(6 v^2(x,t)\  u_x(x,t) + 3 v(x,t)\ w(x,t)\ u_x(x,t) \right) = 0, \\
&& w_t(x,t) + w_{xxx}(x,t) + \left(6 w^2(x,t)\  w_x(x,t) + 3 u(x,t)\  w(x,t)\ v_x(x,t) \right) = 0.  \nonumber
\end{eqnarray}
This is a non-symmetric, nonlinear, coupled mKdV-type system. The system reduces to integrable scalar mKdV equation for
$ u = v = w $. Hence the above system is not integrable in the sense of soliton theory, unless it is symmetrized or reduced to a scalar mKdV equation. Under the following transformation 
\begin{eqnarray}\label{ex2eq2}
u(x,t)=U(\xi), \ \ v(x,t)=V(\xi), \ \ w(x,t)=W(\xi), \ \ \xi=  \alpha \ x +  \beta \  t,
\end{eqnarray}
 equation (\ref{ex2eq1}) reduces to the following system of differential equations
 \begin {eqnarray}\label{ex2eq3}
U^{(3)}(\xi )-\sigma _1^2\  U'(\xi )+\frac{1}{\alpha ^2} \left(6 U(\xi )^2 U'(\xi )+3 U(\xi ) V(\xi ) W'(\xi )\right)=0, \nonumber \\
V^{(3)}(\xi )-\sigma _1^2\  V'(\xi )+\frac{1}{\alpha ^2} \left(6 V(\xi )^2 U'(\xi )+3 V(\xi ) W(\xi ) U'(\xi )\right)=0,\\
W^{(3)}(\xi )-\sigma _1^2\  W'(\xi )+\frac{1}{\alpha ^2} \left(6 W(\xi )^2 W'(\xi )+3 U(\xi ) W(\xi ) V'(\xi )\right)=0,  \nonumber
\end{eqnarray}
 where $\sigma _1=\sqrt{- \frac{\beta}{\alpha^3}}$. It can be reduce to the following matrix form 
 \begin {eqnarray}\label{ex2eq4}
{\cal P}_0(\xi).\ {\bf y }^{(3)}(\xi) - {\cal P}_1(\xi)\ {\bf y}^{(1)}(\xi)  = {\cal N}[{\bf y}](\xi), \, \, \, \, \xi \in \mathbb{R},
\end{eqnarray}
where
\begin {eqnarray}\label{ex2eq5}
 && {\cal P}_0(\xi)=\left[\begin{array}{cccc}
   1&   0&0  \\  0&   1 &0   \\    0 &   0 & 1  \\ \end{array}\right], \ {\cal P}_1(\xi)=\left[\begin{array}{cccc}
   \sigma _1^2&   0&0  \\  0&   \sigma _1^2 &0   \\    0 &   0 & \sigma _1^2  \\\end{array}\right], \  {\bf y }(\xi)=\left[ \begin{array}{c}
   U(\xi )  \\   V(\xi )  \\  W(\xi ) \\ \end{array}\right],  \\  && \ \ \text{and} \ \  {\cal N}[{\bf y}](\xi)=\left[ \begin{array}{c}
   -\frac{1}{\alpha ^2} \left(6 U(\xi )^2 U'(\xi )+3 U(\xi ) V(\xi ) W'(\xi )\right)  \\    -\frac{1}{\alpha ^2} \left(6 V(\xi )^2 U'(\xi )+3 V(\xi ) W(\xi ) U'(\xi )\right)  \\    -\frac{1}{\alpha ^2} \left(6 W(\xi )^2 W'(\xi )+3 U(\xi ) W(\xi ) V'(\xi )\right) \\\end{array}\right]. 
\end{eqnarray}
 Next, we can reduce the equation (\ref{ex2eq4}) to the linear system of equations (\ref{eq1p7}) by following the section \ref{red}.    The linear part of these equations i.e.,  $\hat{\cal L}\left[{\bf y}(\xi) \right]  \simeq {\cal P}_0(\xi).\ {\bf y }^{(3)}(\xi) + {\cal P}_1(\xi).\ {\bf y}^{(1)}(\xi) = 0$ has solution in this form: 
 \begin {eqnarray}\label{ex2eq6}
{\bf y }(\xi)= \tilde{{\bf y }}_1(\xi). \tilde{c}_1 + \tilde{{\bf y }}_2(\xi). \tilde{c}_2 + \tilde{{\bf y }}_3(\xi). \tilde{c}_3
\end{eqnarray} 
where
\begin {eqnarray}\label{ex2eq7}
&& \tilde{{\bf y }}_1(\xi)= e^{-\sqrt{{\cal P}_1(\xi)} \xi } =  \left[
\begin{array}{ccc}
 e^{- \sigma _1 \ \xi} & 0 & 0 \\
 0 &  e^{- \sigma _1 \ \xi} & 0 \\
 0 & 0 &  e^{- \sigma _1 \ \xi} \\
\end{array}
\right],\ \tilde{{\bf y }}_2(\xi)=  e^{\sqrt{{\cal P}_1(\xi)} \xi }=  \left[
\begin{array}{ccc}
  e^{ \sigma _1 \ \xi} & 0 & 0 \\
 0 &  e^{ \sigma _1 \ \xi} & 0 \\
 0 & 0 &  e^{ \sigma _1 \ \xi} \\
\end{array}
\right], \nonumber \\  && \ \ \ \ \ \ \ \ \  
 \  \tilde{{\bf y }}_3(\xi) =I=   \left[
\begin{array}{ccc}
 1 & 0 & 0 \\
 0 & 1 & 0 \\
 0 & 0 & 1 \\
\end{array}
\right],\    \ \ \text{and} \ \
\tilde{c}_i = \left[ \begin{array}{c}
   c_{i,1}  \\    c_{i,2}  \\    c_{i,3} \\\end{array}\right],\ i=1,2,3.
\end{eqnarray} 
The fundamental matrix solutions are
\begin {eqnarray}\label{ex2eq8}
\left\{ \tilde{{\bf y }}_1(\xi)= e^{-\sqrt{{\cal P}_1(\xi)} \xi },\ \tilde{{\bf y }}_2(\xi)= e^{\sqrt{{\cal P}_1(\xi)} \xi }, \  \tilde{{\bf y }}_3(\xi)=I \right\},
\end{eqnarray} 
where $I$ is an identity matrix of size $3\times 3$ and ${\cal P}_1(\xi)$ is defined as previously. Now, by changing the constant matrices $\tilde{c}_i$ to functions of $\xi$ and applying (\ref{eq2p13}), MMVP can solve this system of equations by a recursive approach
 \begin {eqnarray}\label{ex2eq9}
 \begin{cases}
  {\bf y}_0(\xi)=\tilde{{\bf y }}_1(\xi). \tilde{c}_1 + \tilde{{\bf y }}_2(\xi). \tilde{c}_2 + \tilde{{\bf y }}_3(\xi). \tilde{c}_3 \left(\equiv \sum_{l=1}^{n} \tilde{{\bf y }}_l(\xi). \tilde{c}_l \right),  \\
 {\bf y}_k(\xi) = \sum_{l=1}^{n} \left\{   \tilde{{\bf y} }_{l}(\xi). \bigintsss \left[ \frac{ 1}{ W(\xi)} \ \Bigg( (-1)^{i+ j} W_{i,j}(\xi)\Bigg)^T _{\substack{i = m n - (m-1),\ \cdots ,\ m n  \\ j = (l-1)m + 1,\ \cdots ,\ l m }}.\left({\cal P}_0(\xi)^{-1}.\ {\cal A}_{k-1}(\xi)\right) \right] d\xi \right\},   \ \ k \geq 1,
\end{cases}
\end{eqnarray}
where ${\cal A}_{k-1}(\xi)$ are matrix of Adomian polynomials of nonlinear matrix term  ${\cal N}[U](\xi)$ calculated using the formula (\ref{eq1p5}),  $W(\xi)$ represents the block Wronskian of the fundamental matrix solution set $\left\{ \tilde{{\bf y }}_1(\xi),\ \tilde{{\bf y }}_2(\xi), \ \tilde{{\bf y }}_3(\xi)\right\}$,  and  $W_{i,j}(\xi)$ are  determinants, derived by removing the $i-$th row and the $j-$th column from $W(\xi)$.\\ 
\textbf{Case-1.\  For Boundary Condition $U(-\infty)= 0,\ V(-\infty)= 0,\ W(-\infty)= 0.$ }\\
Using the boundary condition $U(\xi)\rightarrow 0,\ V(\xi)\rightarrow 0,\ W(\xi)\rightarrow 0$ as $\xi \rightarrow - \infty $ for localized solution on recursive scheme (\ref{ex2eq4}), we now obtain $c_{1,i}=c_{3,i}=0,\ i=1,2,3$ which makes $${\bf y}_0(\xi)=\left[
\begin{array}{c}
 c_{2,1} \  e^{ \sigma _1 \ \xi}\\
  c_{2,2} \  e^{ \sigma _1 \ \xi} \\
  c_{2,3} \  e^{ \sigma _1 \ \xi} \\
\end{array}
\right],$$ when $\sigma_1 >0$. The following correction terms are obtained using this modified ${\bf y}_0(\xi)$ and iterative technique (\ref{ex2eq4}).
\begin {eqnarray}\label{ex2eq10}
&& {\bf y}_0(\xi)=\left[ \begin{array}{c}
   U_0(\xi )  \\   V_0(\xi )  \\  W_0(\xi ) \\ \end{array}\right]=\left[
\begin{array}{c}
 c_{2,1} \  e^{ \sigma _1 \ \xi}\\
  c_{2,2} \  e^{ \sigma _1 \ \xi} \\
  c_{2,3} \  e^{ \sigma _1 \ \xi} \\
\end{array}
\right],\ {\bf y}_1(\xi)=\left[ \begin{array}{c}
   U_1(\xi )  \\   V_1(\xi )  \\  W_1(\xi ) \\ \end{array}\right]=\left[
\begin{array}{c}
 -\frac{c_{2,1} \left(2
   c_{2,1}^2+c_{2,2} c_{2,3}\right)  }{8 \alpha ^2 \sigma
   _1^2} \ e^{3 \  \sigma _1 \ \xi} \\
 -\frac{ c_{2,1} c_{2,2} \left(2
   c_{2,2}+c_{2,3}\right)  }{8 \alpha ^2 \sigma _1^2} \ e^{3 \  \sigma _1 \ \xi}  \\
 -\frac{c_{2,3} \left(2
   c_{2,3}^2+c_{2,1} c_{2,2}\right) }{8 \alpha ^2 \sigma
   _1^2}  \ e^{3 \  \sigma _1 \ \xi} \\
\end{array}
\right],\\ 
   && {\bf y}_2(\xi)=\left[ \begin{array}{c}
   U_2(\xi )  \\   V_2(\xi )  \\  W_2(\xi ) \\ \end{array}\right]=\left[
\begin{array}{c}
 \frac{ c_{2,1} \left\{20 c_{2,1}^4+12
   c_{2,2} c_{2,3} c_{2,1}^2+c_{2,2} c_{2,3} \left(5
   c_{2,2}+c_{2,3}\right) c_{2,1}+c_{2,2} c_{2,3}^2
   \left(c_{2,2}+6 c_{2,3}\right)\right\} }{320 \alpha ^4
   \sigma _1^4}  \ e^{5 \  \sigma _1 \ \xi} \\
 \frac{ c_{2,1} c_{2,2} \left\{6 \left(2
   c_{2,2}+c_{2,3}\right) c_{2,1}^2+\left(8 c_{2,2}^2+7
   c_{2,3} c_{2,2}+c_{2,3}^2\right) c_{2,1}+c_{2,3} \left(6
   c_{2,2}^2+3 c_{2,3} c_{2,2}+2
   c_{2,3}^2\right)\right\} }{320 \alpha ^4 \sigma _1^4}  \ e^{5 \  \sigma _1 \ \xi} \\
 \frac{ c_{2,3} \left\{20
   c_{2,3}^4+c_{2,1} c_{2,2} \left(c_{2,2}+12
   c_{2,3}\right) c_{2,3}+2 c_{2,1}^3 c_{2,2}+c_{2,1}^2
   c_{2,2} \left(7 c_{2,2}+3 c_{2,3}\right)\right\} }{320
   \alpha ^4 \sigma _1^4}  \ e^{5 \  \sigma _1 \ \xi} \\
\end{array}
\right], \\
&& \tiny{ {\bf y}_3(\xi)=\left[ \begin{array}{c}
   U_3(\xi )  \\   V_3(\xi )  \\  W_3(\xi ) \\ \end{array}\right]=\left[
\begin{array}{c}
 -  e^{7 \xi  \sigma _1} c_{2,1} \left\{560
   c_{2,1}^6+468 c_{2,2} c_{2,3} c_{2,1}^4+2 c_{2,2}
   c_{2,3} \left(71 c_{2,2}+15 c_{2,3}\right)
   c_{2,1}^3+c_{2,2} c_{2,3} \left(73 c_{2,2}^2+133 c_{2,3}
   c_{2,2}+145 c_{2,3}^2\right) c_{2,1}^2 \right. \\ \left. +c_{2,2} c_{2,3}^2
   \left(41 c_{2,2}^2+129 c_{2,3} c_{2,2}+32
   c_{2,3}^2\right) c_{2,1}+c_{2,2} c_{2,3}^3
   \left(c_{2,2}^2+36 c_{2,3} c_{2,2}+100
   c_{2,3}^2\right)\right\} \mathbin{/}( 35840 \alpha ^6 \sigma _1^6) \\
 - e^{7 \xi  \sigma _1} c_{2,1} c_{2,2} \left\{100
   \left(2 c_{2,2}+c_{2,3}\right) c_{2,1}^4+4 \left(72
   c_{2,2}^2+62 c_{2,3} c_{2,2}+9 c_{2,3}^2\right)
   c_{2,1}^3+\left(72 c_{2,2}^3+213 c_{2,3} c_{2,2}^2+89
   c_{2,3}^2 c_{2,2}+61 c_{2,3}^3\right) c_{2,1}^2  \right. \\ \left. +c_{2,3}
   \left(194 c_{2,2}^3+159 c_{2,3} c_{2,2}^2+63 c_{2,3}^2
   c_{2,2}+12 c_{2,3}^3\right) c_{2,1}+5 c_{2,3}^2 \left(2
   c_{2,2}^3+13 c_{2,3} c_{2,2}^2+12 c_{2,3}^2 c_{2,2}+4
   c_{2,3}^3\right)\right\} \mathbin{/}({35840 \alpha ^6 \sigma _1^6}) \\
 - e^{7 \xi  \sigma _1} c_{2,3} \left\{560
   c_{2,3}^6+c_{2,1} c_{2,2} \left(c_{2,2}^2+30 c_{2,3}
   c_{2,2}+468 c_{2,3}^2\right) c_{2,3}^2+c_{2,1}^2 c_{2,2}
   \left(71 c_{2,2}^2+271 c_{2,3} c_{2,2}+82
   c_{2,3}^2\right) c_{2,3}+20 c_{2,1}^5 c_{2,2}  \right. \\ \left. +12
   c_{2,1}^4 c_{2,2} \left(11 c_{2,2}+5
   c_{2,3}\right)+c_{2,1}^3 c_{2,2} \left(77 c_{2,2}^2+65
   c_{2,3} c_{2,2}+53 c_{2,3}^2\right)\right\} \mathbin{/}(35840 \alpha
   ^6 \sigma _1^6) \\
\end{array}
\right], } \\
&&  \cdots 
\end{eqnarray}
Likewise, symbolic calculations can be used to determine higher order correction terms. 
Thus, the series solution mentioned above can be got into the form
\begin {eqnarray}\label{ex2eq11}
&&{\bf y }(\xi)=\left[ \begin{array}{c}
   U(\xi )  \\   V(\xi )  \\  W(\xi ) \\ \end{array}\right]=\sum_{n=0}^\infty \left[ \begin{array}{c}
   U_n(\xi )  \\   V_n(\xi )  \\  W_n(\xi ) \\ \end{array}\right]= \left[ \begin{array}{c}
   \sum_{n=0}^\infty U_n(\xi )  \\   \sum_{n=0}^\infty V_n(\xi )  \\  \sum_{n=0}^\infty W_n(\xi ) \\ \end{array}\right] \nonumber \\
&& =\left[ \begin{array}{c}
   c_{2,1} \  e^{ \sigma _1 \ \xi} -\frac{c_{2,1} \left(2
   c_{2,1}^2+c_{2,2} c_{2,3}\right)  }{8 \alpha ^2 \sigma
   _1^2} \ e^{3 \  \sigma _1 \ \xi}+ \frac{ c_{2,1} \left\{20 c_{2,1}^4+12
   c_{2,2} c_{2,3} c_{2,1}^2+c_{2,2} c_{2,3} \left(5
   c_{2,2}+c_{2,3}\right) c_{2,1}+c_{2,2} c_{2,3}^2
   \left(c_{2,2}+6 c_{2,3}\right)\right\} }{320 \alpha ^4
   \sigma _1^4}  \ e^{5 \  \sigma _1 \ \xi} +\cdots  \\  c_{2,2} \  e^{ \sigma _1 \ \xi}-\frac{ c_{2,1} c_{2,2} \left(2
   c_{2,2}+c_{2,3}\right)  }{8 \alpha ^2 \sigma _1^2} \ e^{3 \  \sigma _1 \ \xi} +\frac{ c_{2,1} c_{2,2} \left\{6 \left(2
   c_{2,2}+c_{2,3}\right) c_{2,1}^2+\left(8 c_{2,2}^2+7
   c_{2,3} c_{2,2}+c_{2,3}^2\right) c_{2,1}+c_{2,3} \left(6
   c_{2,2}^2+3 c_{2,3} c_{2,2}+2
   c_{2,3}^2\right)\right\} }{320 \alpha ^4 \sigma _1^4}  \ e^{5 \  \sigma _1 \ \xi}+\cdots \\   c_{2,3} \  e^{ \sigma _1 \ \xi} -\frac{c_{2,3} \left(2
   c_{2,3}^2+c_{2,1} c_{2,2}\right) }{8 \alpha ^2 \sigma
   _1^2}  \ e^{3 \  \sigma _1 \ \xi}+ \frac{ c_{2,3} \left\{20
   c_{2,3}^4+c_{2,1} c_{2,2} \left(c_{2,2}+12
   c_{2,3}\right) c_{2,3}+2 c_{2,1}^3 c_{2,2}+c_{2,1}^2
   c_{2,2} \left(7 c_{2,2}+3 c_{2,3}\right)\right\} }{320
   \alpha ^4 \sigma _1^4}  \ e^{5 \  \sigma _1 \ \xi} +\cdots \\ \end{array}\right].  \nonumber \\
\end{eqnarray}
 Since the consider equation is not integrable, the closed form of the above derived series $ \left(U(\xi )=\sum_{n=0}^\infty U_n(\xi ),\right. $ $ \left. V(\xi )=\sum_{n=0}^\infty V_n(\xi ),\ W(\xi )=\sum_{n=0}^\infty W_n(\xi ) \right)$ can not be obtained for general values of the involve parameters. We employ the multiplicative inverse of power series method described in the section \ref{mi} to obtain the closed form of the derived series for specific values of the involved parameters.
\\ 
\textbf{ $\blacksquare$  Multiplicative inverse of power series (\ref{ex2eq11}) and exact solution: }   
After the substitution $ e^{ \sigma _1 \ \xi} = \zeta $, the series $ \left(U(\xi )=\sum_{n=0}^\infty U_n(\xi ),\right. $ $ \left. V(\xi )=\sum_{n=0}^\infty V_n(\xi ),\ W(\xi )=\sum_{n=0}^\infty W_n(\xi ) \right)$ can be put in terms of $\zeta$ in the  form $\left( U(\zeta )=  \sum_{n=0}^\infty U_n(\zeta),\right. $ $ \left. V(\zeta )=\sum_{n=0}^\infty V_n(\zeta),\ W(\zeta)=\sum_{n=0}^\infty W_n(\zeta) \right)$. Since the existence of multiplicative inverse of a power series demand the first constant term non-zero, we make a small adjustment instead of original power series we find the multiplicative inverse of the  series $\left(\frac{U(\zeta )}{ \zeta } \ ,\right. $ $ \left. \frac{V(\zeta )}{ \zeta } ,\ \frac{W(\zeta )}{ \zeta } \right) $.  
To obtain the corresponding inverse power series, we use the recursive formula (\ref{mi1}) for deriving multiplicative inverse series as follows:
\begin {eqnarray}\label{mip1}
&&\left( \frac{U(\zeta )}{ \zeta }\right)^{-1}=  \sum_{n=0}^\infty \tilde{U}_n(\zeta) = \frac{1}{c_{2,1}}+\frac{ \left(2 c_{2,1}^2+c_{2,2} c_{2,3}\right)}{8 \alpha ^2
   \sigma _1^2 c_{2,1}} \zeta ^2+\frac{ c_{2,2} c_{2,3} \left(8 c_{2,1}^2-5 c_{2,2}
   c_{2,1}-c_{2,3} c_{2,1}-6 c_{2,3}^2+4 c_{2,2} c_{2,3}\right)}{320
   \alpha ^4 \sigma _1^4 c_{2,1}} \zeta ^4+\cdots \nonumber \\
&& \left( \frac{V(\zeta )}{ \zeta }\right)^{-1} =  \sum_{n=0}^\infty \tilde{V}_n(\zeta)= \frac{1}{c_{2,2}}+\frac{c_{2,1} \left(2 c_{2,2}+c_{2,3}\right)}{8 \alpha ^2 \sigma _1^2
   c_{2,2}} \zeta ^2 \nonumber \\
&& \ \ \ \   -\frac{c_{2,1} \left\{6 \left(2 c_{2,2}+c_{2,3}\right) c_{2,1}^2-\left(12
   c_{2,2}^2+13 c_{2,3} c_{2,2}+4 c_{2,3}^2\right) c_{2,1}+c_{2,3}
   \left(6 c_{2,2}^2+3 c_{2,3} c_{2,2}+2 c_{2,3}^2\right)\right\}}{320
   \alpha ^4 \sigma _1^4 c_{2,2}} \zeta ^4+\cdots  \\
&& \left( \frac{W(\zeta )}{ \zeta }\right)^{-1}=  \sum_{n=0}^\infty \tilde{W}_n(\zeta) = \frac{1}{c_{2,3}}+\frac{2 c_{2,3}^2+c_{2,1} c_{2,2}}{8 \alpha ^2 \sigma _1^2 c_{2,3}} \zeta ^2+\frac{c_{2,1} c_{2,2} \left(-2 c_{2,1}^2-2 c_{2,2} c_{2,1}-3 c_{2,3}
   c_{2,1}+8 c_{2,3}^2-c_{2,2} c_{2,3}\right)}{320 \alpha ^4 \sigma _1^4
   c_{2,3}} \zeta ^4+\cdots \nonumber 
\end{eqnarray} 
Similarly we can calculate the higher order recurrence terms utilizing symbolic computations. To find the closed form or generating function of the above series we follow section \ref{mi}. 1) Truncating  the series: We need to terminate the above infinite series after finite term by equating the coefficients of $\zeta $ after certain term to zero. The first terms of above series are non-zero. Equating the second terms (i.e., coefficients of  $\zeta ^2$) of above three series to zero leads us to trivial solution. Hence equating the coefficients of  $\zeta ^4$ to zero and solving this system of equation we get the following solution:
\begin {eqnarray}\label{mip2}
&& \text{Set-1:} \ \ \ \ c_{2,1}= -2\ c_{2,3},\ c_{2,2}= -2\  c_{2,3}, \\
&& \text{Set-2:}   \ \ \ \ c_{2,1}= \frac{c_{2,3}}{2},\ c_{2,2}= 3\ c_{2,3}, \\
&& \text{Set-3:}  \ \ \ \    c_{2,1}= c_{2,3},\  c_{2,2}= c_{2,3}. 
\end{eqnarray}
Notably, the aforementioned three series end after finite terms for the restrictions mentioned above.\\
2)  Summing infinite series: To find the generating function of these infinite series we comparing it with a convergent series. We assume that the corresponding series higher terms have a common ration i.e., 
$$\frac{\tilde{U}_{n+1}(\zeta)}{\tilde{U}_n(\zeta)}=\frac{\tilde{U}_{n+2}(\zeta)}{\tilde{U}_{n+1}(\zeta)}, \ \ \ \frac{\tilde{V}_{n+1}(\zeta)}{\tilde{V}_n(\zeta)}=\frac{\tilde{V}_{n+2}(\zeta)}{\tilde{V}_{n+1}(\zeta)},\ \ \ \frac{\tilde{W}_{n+1}(\zeta)}{\tilde{W}_n(\zeta)}=\frac{\tilde{W}_{n+2}(\zeta)}{\tilde{W}_{n+1}(\zeta)}, \ \ n\geq 1. $$ We are unable to find any suitable solution for the above equations for any value of $n$.  Therefore, multiplicative inverse does not offer any appropriate answers for the aforementioned constraints. Nevertheless, if we apply the aforementioned constraints to the original series (\ref{ex2eq11}) in place of multiplicative inverse power series (\ref{mip1}), we will obtain a set of appropriate solutions same as provided in the case of truncating series.    In the subsequent sections, we will use the conditions to determine closed form solutions of (\ref{ex2eq6}) or $ \left(U(\xi ),\right. $ $ \left. V(\xi ),\ W(\xi ) \right)$.\\
\textbf{Solution-1: \ } We now obtain the close form solution of the considered series by applying the restriction \text{Set-1} in the following form:
\begin {eqnarray}\label{mip3}
&&{\bf y }(\zeta)=\left[ \begin{array}{c}
   U(\zeta )  \\   V(\zeta)  \\  W(\zeta) \\ \end{array}\right]=\left[ \begin{array}{c}
  \frac{\zeta}{ \left( \frac{U(\zeta )}{ \zeta }\right)^{-1}}  \\    \frac{\zeta}{ \left( \frac{V(\zeta )}{ \zeta }\right)^{-1}}   \\  \frac{\zeta}{ \left( \frac{W(\zeta )}{ \zeta }\right)^{-1}}  \\ \end{array}\right]=\left[ \begin{array}{c}
 \frac{
  8 \alpha^{2} \sigma_1^{2} c_{2,3} \, \zeta
}{
  -4 \alpha^{2} \sigma_1^{2} - 3 c_{2,3}^2 \ \zeta^2
}

  \\   
\frac{
8 \alpha^{2} \sigma_1^{2} c_{2,3} \, \zeta
}{
- \left(4 \alpha^{2} \sigma_1^{2} + 3 c_{2,3}^2 \ \zeta^2 \right)
}
 \\ 
 \frac{4 \alpha^{2} \sigma_1^{2} c_{2,3} \, \zeta }{4 \alpha^{2} \sigma_1^{2} + 3 c_{2,3}^2 \ \zeta^2 }
 \\ \end{array}\right].
\end{eqnarray}
The considered series (\ref{ex2eq6}) has the following close form solution when one returns to the original variable $\xi$.
\begin {eqnarray}\label{mip4}
&&{\bf y }(\xi)=\left[ \begin{array}{c}
   U(\xi )  \\   V(\xi)  \\  W(\xi) \\ \end{array}\right]=\left[ \begin{array}{c}
 \frac{
  8 \alpha^{2} \sigma_1^{2} c_{2,3} \, e^{\xi \sigma_1}
}{
  -4 \alpha^{2} \sigma_1^{2} - 3 c_{2,3}^2 e^{2 \xi \sigma_1}
}

  \\   
\frac{
8 \alpha^{2} \sigma_1^{2} c_{2,3} \, e^{\xi \sigma_1}
}{
- \left(4 \alpha^{2} \sigma_1^{2} + 3 c_{2,3}^2 e^{2 \xi \sigma_1}\right)
}
 \\ 
 \frac{4 \alpha^{2} \sigma_1^{2} c_{2,3} \, e^{\xi \sigma_1}}{4 \alpha^{2} \sigma_1^{2} + 3 c_{2,3}^2 e^{2 \xi \sigma_1}}
 \\ \end{array}\right].
\end{eqnarray}
\textbf{Solution-2: \ } Following the procedures in Solution-1 and applying the restriction \text{Set-2}, we obtain the close form solution of the series under consideration (\ref{ex2eq6}) in the form
\begin {eqnarray}\label{mip5}
&&{\bf y }(\xi)=\left[ \begin{array}{c}
   U(\xi )  \\   V(\xi)  \\  W(\xi) \\ \end{array}\right]=\left[ \begin{array}{c}

\frac{
8 \alpha^{2} \sigma_1^{2} c_{2,3} \, e^{\xi \sigma_1}
}{
16 \alpha^{2} \sigma_1^{2} + 7 c_{2,3}^2 e^{2 \xi \sigma_1}
}

  \\   

\frac{
144 \alpha^{2} \sigma_1^{2} c_{2,3} \, e^{\xi \sigma_1}
}{
48 \alpha^{2} \sigma_1^{2} + 21 c_{2,3}^2 e^{2 \xi \sigma_1}
}
 \\ 

\frac{16 \alpha^{2} \sigma_1^{2} c_{2,3} \, e^{\xi \sigma_1}}{16 \alpha^{2} \sigma_1^{2} + 7 c_{2,3}^2 e^{2 \xi \sigma_1}}
 \\ \end{array}\right].
\end{eqnarray}
\textbf{Solution-3: \ } Following the procedures in case Solution-1 and applying the restriction \text{Set-3}, we obtain another close form solution of the considered series (\ref{ex2eq6}) in the form
\begin {eqnarray}\label{mip6}
&&{\bf y }(\xi)=\left[ \begin{array}{c}
   U(\xi )  \\   V(\xi)  \\  W(\xi) \\ \end{array}\right]=\left[ \begin{array}{c}

\frac{8 \alpha^{2} \sigma_1^{2} c_{2,3} \, e^{\xi \sigma_1}}{8 \alpha^{2} \sigma_1^{2} + 3 c_{2,3}^{2} e^{2 \xi \sigma_1}}

  \\   

\frac{8 \alpha^{2} \sigma_1^{2} c_{2,3} \, e^{\xi \sigma_1}}{8 \alpha^{2} \sigma_1^{2} + 3 c_{2,3}^2 e^{2 \xi \sigma_1}}
 \\ 

\frac{8 \alpha^{2} \sigma_1^{2} c_{2,3} \, e^{\xi \sigma_1}}{8 \alpha^{2} \sigma_1^{2} + 3 c_{2,3}^2 e^{2 \xi \sigma_1}}
 \\ \end{array}\right].
\end{eqnarray}
Notably, (\ref{mip4}), (\ref{mip5}), and (\ref{mip6}) with (\ref{ex2eq2}) yield the three sets of solutions of (\ref{ex2eq1}).\\
\textbf{Case-2.\  For Boundary Condition $U(\infty)= 0,\ V(\infty)= 0,\ W(\infty)= 0.$ }\\
Once the boundary condition $U(\xi)\rightarrow 0,\ V(\xi)\rightarrow 0,\ W(\xi)\rightarrow 0$ is applied, $c_{2,i}=c_{3,i}=0,\ i=1,2,3$ is obtained as $\xi \rightarrow \infty $ for localized solution on recursive scheme (\ref{ex2eq9}). We may get a solution that is similar to (\ref{mip4}), (\ref{mip5}), and (\ref{mip6}) using these values and the procedures of Case-1, except for the change in sign of the independent variable. For the sake of simplicity, we excluded these calculations out of the study. 
\subsection{Example 2:} In this example, by using MMVP we will study the solution structures of the following integrable coupled KdV equation \citep{cao2010prolongation}
\begin{eqnarray}\label{ex2ex1}
&& u_t(x,t) -\alpha_1 \ \left(-u_{xxx}(x,t) + 6 u(x,t) u_x(x,t) + 3 v(x,t)\  w_x(x,t) + 3 w(x,t)\  v_x(x,t)\right) = 0, \nonumber
 \\
&& v_t(x,t) -\alpha_1 \ \left(- v_{xxx}(x,t) + 6 u(x,t) v_x(x,t) + 6 v(x,t)\ u_x(x,t)  \right) = 0, \\
&& w_t(x,t) -\alpha_1 \ \left(- w_{xxx}(x,t) + 6 u(x,t) w_x(x,t) + 6 w(x,t)\ u_x(x,t) \right) = 0.  \nonumber
\end{eqnarray} 
These coupled KdV equations have many applications in several physical fields, such as shallow stratified liquid \citep{gear1984weak}, atmospheric dynamical system \citep{cao2010prolongation} and so on.  It should be noted that under the following Miura transformation
\begin{eqnarray}\label{ex2ex1a}
&& u(x,t)= p_x(x,t)+p(x,t)^2+q(x,t)\ r(x,t),\\\nonumber
&& v(x,t)=  2 p(x,t)\ q(x,t)+q_x(x,t),\\
&& w(x,t)= 2 p(x,t)\ r(x,t)+r_x(x,t), \nonumber
\end{eqnarray}
equation (\ref{ex2ex1})  reduced to the following new integrable coupled modified KdV equations
\begin{eqnarray}\label{ex2ex1b}
&& p_t(x,t) -\alpha_1 \ \left(- p_{xxx}(x,t) + 6 p^2(x,t) p_x(x,t) + 6 p(x,t)\ q(x,t) \  r_x(x,t) + 6 p(x,t)\ r(x,t)\ q_x(x,t) \right. \nonumber \\ && \ \ \ \ \ \ \ \ \ \ \ \ \ \ \left.+ 6 r(x,t)\ q(x,t)\ p_x(x,t)  \right) = 0, \nonumber
 \\
&& q_t(x,t) -\alpha_1 \ \left(- q_{xxx}(x,t) + 6 p^2(x,t) q_x(x,t) + 12 p(x,t)\ q(x,t) \  p_x(x,t) + 6 q(x,t)\ r(x,t)\ q_x(x,t)  \right) = 0, \\
&& r_t(x,t) -\alpha_1 \ \left(- r_{xxx}(x,t) + 6 p^2(x,t) r_x(x,t) + 12 p(x,t)\ r(x,t) \  p_x(x,t) + 6 q(x,t)\ r(x,t)\ r_x(x,t) \right) = 0.  \nonumber
\end{eqnarray}
The above system is  integrable in the sense of soliton theory. Under the following transformation 
\begin{eqnarray}\label{ex2ex2}
p(x,t)=U(\xi), \ \ q(x,t)=V(\xi), \ \ r(x,t)=W(\xi), \ \ \xi= k ( x -  c \  t),
\end{eqnarray}
 the equation (\ref{ex2ex1b}) reduces to the following system of differential equations
 \begin {eqnarray}\label{ex2ex3}
U^{(3)}(\xi )-\lambda^2\  U'(\xi )-\frac{6}{k^2} \left\{ U(\xi )^2 U'(\xi )+\left( U(\xi ) V(\xi ) W(\xi ) \right)' \right\}=0, \nonumber \\
V^{(3)}(\xi )-\lambda^2\  V'(\xi )-\frac{6}{k^2} \left\{ V(\xi ) W(\xi ) V'(\xi )+ \left(U(\xi )^2 V(\xi )\right)'\right\}=0,\\
W^{(3)}(\xi )-\lambda^2\  W'(\xi )-\frac{6}{k^2}  \left\{ V(\xi ) W(\xi ) W'(\xi )+ \left(U(\xi )^2 W(\xi )\right)'\right\}=0,  \nonumber
\end{eqnarray}
 where $\lambda=\sqrt{\frac{c}{\alpha _1 k^2}}$. It can be reduce to the following matrix form 
 \begin {eqnarray}\label{ex2ex4}
{\cal P}_0(\xi).\ {\bf y }^{(3)}(\xi) - {\cal P}_1(\xi)\ {\bf y}^{(1)}(\xi)  = {\cal N}[{\bf y}](\xi), \, \, \, \, \xi \in \mathbb{R},
\end{eqnarray}
where
\begin {eqnarray}\label{ex2ex5}
&& {\cal P}_0(\xi)=\left[\begin{array}{cccc}
   1&   0&0  \\  0&   1 &0   \\    0 &   0 & 1  \\ \end{array}\right], \ {\cal P}_1(\xi)=\left[\begin{array}{cccc}
   \lambda^2&   0&0  \\  0&  \lambda^2 &0   \\    0 &   0 & \lambda^2  \\\end{array}\right], \  {\bf y }(\xi)=\left[ \begin{array}{c}
   U(\xi )  \\   V(\xi )  \\  W(\xi ) \\ \end{array}\right],  \\ && \ \ \ \ \ \text{and}  \ \ \  {\cal N}[{\bf y}](\xi)=\left[ \begin{array}{c}
  \frac{6}{k^2} \left\{ U(\xi )^2 U'(\xi )+\left( U(\xi ) V(\xi ) W(\xi ) \right)' \right\}  \\    \frac{6}{k^2}  \left\{ V(\xi ) W(\xi ) V'(\xi )+ \left(U(\xi )^2 V(\xi )\right)'\right\}  \\   \frac{6}{k^2}   \left\{ V(\xi ) W(\xi ) W'(\xi )+ \left(U(\xi )^2 W(\xi )\right)'\right\} \\\end{array}\right]. 
\end{eqnarray}
 Next following the section \ref{red}, we can reduce the equation (\ref{ex2ex4}) to the linear system of equation (\ref{eq1p7}).   The linear part of these equations i.e.,  $\hat{\cal L}\left[{\bf y}(\xi) \right]  \simeq {\cal P}_0(\xi).\ {\bf y }^{(3)}(\xi) + {\cal P}_1(\xi).\ {\bf y}^{(1)}(\xi) = 0$ has solution in this form: 
 \begin {eqnarray}\label{ex2ex6}
{\bf y }(\xi)= \tilde{{\bf y }}_1(\xi). \tilde{c}_1 + \tilde{{\bf y }}_2(\xi). \tilde{c}_2 + \tilde{{\bf y }}_3(\xi). \tilde{c}_3
\end{eqnarray} 
where
\begin {eqnarray}\label{ex2ex7}
&& \tilde{{\bf y }}_1(\xi)= e^{-\sqrt{{\cal P}_1(\xi)} \xi } =  \left[
\begin{array}{ccc}
 e^{- \lambda \ \xi} & 0 & 0 \\
 0 &  e^{- \lambda \ \xi} & 0 \\
 0 & 0 & e^{- \lambda \ \xi} \\
\end{array}
\right],\ \tilde{{\bf y }}_2(\xi)=  e^{\sqrt{{\cal P}_1(\xi)} \xi }=  \left[
\begin{array}{ccc}
  e^{ \lambda \ \xi} & 0 & 0 \\
 0 &  e^{ \lambda \ \xi} & 0 \\
 0 & 0 &  e^{ \lambda \ \xi} \\
\end{array}
\right], \nonumber \\ &&  \ \ \   \ \ \   \ \ \ 
 \  \tilde{{\bf y }}_3(\xi) =I=   \left[
\begin{array}{ccc}
 1 & 0 & 0 \\
 0 & 1 & 0 \\
 0 & 0 & 1 \\
\end{array}
\right],\  \ \ \ \text{and}  \ \ \  
\tilde{c}_i = \left[ \begin{array}{c}
   c_{i,1}  \\    c_{i,2}  \\    c_{i,3} \\\end{array}\right],\ i=1,2,3.
\end{eqnarray} 
The fundamental matrix solutions are
\begin {eqnarray}\label{ex2ex8}
\left\{ \tilde{{\bf y }}_1(\xi)= e^{-\sqrt{{\cal P}_1(\xi)} \xi },\ \tilde{{\bf y }}_2(\xi)= e^{\sqrt{{\cal P}_1(\xi)} \xi }, \  \tilde{{\bf y }}_3(\xi)=I \right\},
\end{eqnarray} 
where $I$ is $3\times 3$ identity matrix and ${\cal P}_1(\xi)$  is defined above. Now varying the constant matrices $\tilde{c}_i$ to functions of $\xi$ and 
   using x (\ref{eq2p13}), MMVP gives  the following recursive scheme
 \begin {eqnarray}\label{ex2ex9}
  \begin{cases}
  {\bf y}_0(\xi)=\tilde{{\bf y }}_1(\xi). \tilde{c}_1 + \tilde{{\bf y }}_2(\xi). \tilde{c}_2 + \tilde{{\bf y }}_3(\xi). \tilde{c}_3 \left(\equiv \sum_{l=1}^{n} \tilde{{\bf y }}_l(\xi). \tilde{c}_l \right),  \\
 {\bf y}_k(\xi) = \sum_{l=1}^{n} \left\{   \tilde{{\bf y} }_{l}(\xi). \bigintsss \left[ \frac{ 1}{ W(\xi)} \ \Bigg( (-1)^{i+ j} W_{i,j}(\xi)\Bigg)^T _{\substack{i = m n - (m-1),\ \cdots ,\ m n  \\ j = (l-1)m + 1,\ \cdots ,\ l m }}.\left({\cal P}_0(\xi)^{-1}.\ {\cal A}_{k-1}(\xi)\right) \right] d\xi \right\},   \ \ k \geq 1,
\end{cases}
\end{eqnarray}
where ${\cal A}_{k-1}(\xi)$ are the matrix of Adomian polynomials of the nonlinear matrix term ${\cal N}[U](\xi)$ calculated using the formula (\ref{eq1p5}), and $W(\xi)$ represents the block Wronskian of the fundamental matrix solution set $\left\{ \tilde{{\bf y }}_1(\xi),\ \tilde{{\bf y }}_2(\xi), \ \tilde{{\bf y }}_3(\xi)\right\}$, and $W_{i,j}(\xi)$ is the determinant derived by removing the $i-$th row and the $j-$th column from $W(\xi)$.\\ 
\textbf{Case-1.\  For Boundary Condition $U(-\infty)= 0,\ V(-\infty)= 0,\ W(-\infty)= 0.$ }\\
Using the boundary condition $U(\xi)\rightarrow 0,\ V(\xi)\rightarrow 0,\ W(\xi)\rightarrow 0$ as $\xi \rightarrow - \infty $ for localized solution on recursive scheme (\ref{ex2ex9}), we now obtain $c_{1,i}=c_{3,i}=0,\ i=1,2,3$ which gives $${\bf y}_0(\xi)=\left[
\begin{array}{c}
 c_{2,1} \  e^{ \lambda \ \xi}\\
  c_{2,2} \  e^{ \lambda \ \xi} \\
  c_{2,3} \  e^{ \lambda \ \xi} \\
\end{array}
\right],$$ when $\lambda >0$. Utilizing this modified ${\bf y}_0(\xi)$ in the iterative technique (\ref{ex2ex9}), we obtain the subsequent corrective terms
\begin {eqnarray}\label{ex2ex10}
&& {\bf y}_0(\xi)=\left[ \begin{array}{c}
   U_0(\xi )  \\   V_0(\xi )  \\  W_0(\xi ) \\ \end{array}\right]=\left[
\begin{array}{c}
 c_{2,1} \  e^{ \lambda \ \xi}\\
  c_{2,2} \  e^{ \lambda \ \xi} \\
  c_{2,3} \  e^{ \lambda \ \xi} \\
\end{array}
\right],\ {\bf y}_1(\xi)=\left[ \begin{array}{c}
   U_1(\xi )  \\   V_1(\xi )  \\  W_1(\xi ) \\ \end{array}\right]=\left[
\begin{array}{c}
 \frac{c_{2,1} \left(c_{2,1}^2+3 c_{2,2} c_{2,3}\right) e^{3 \lambda  \xi }}{4 k^2 \lambda ^2} \\
 \frac{c_{2,2} \left(3 c_{2,1}^2+c_{2,2} c_{2,3}\right) e^{3 \lambda  \xi }}{4 k^2 \lambda ^2}  \\
\frac{c_{2,3} \left(3 c_{2,1}^2+c_{2,2} c_{2,3}\right) e^{3 \lambda  \xi }}{4 k^2 \lambda ^2} \\
\end{array}
\right],\\ 
   && {\bf y}_2(\xi)=\left[ \begin{array}{c}
   U_2(\xi )  \\   V_2(\xi )  \\  W_2(\xi ) \\ \end{array}\right]=\left[
\begin{array}{c}
 \frac{c_{2,1} \left(c_{2,1}^4+10 c_{2,2} c_{2,3} c_{2,1}^2+5 c_{2,2}^2 c_{2,3}^2\right) e^{5 \lambda  \xi }}{16 k^4 \lambda ^4} \\
 \frac{c_{2,2} \left(5 c_{2,1}^4+10 c_{2,2} c_{2,3} c_{2,1}^2+c_{2,2}^2 c_{2,3}^2\right) e^{5 \lambda  \xi }}{16 k^4 \lambda ^4} \\
\frac{c_{2,3} \left(5 c_{2,1}^4+10 c_{2,2} c_{2,3} c_{2,1}^2+c_{2,2}^2 c_{2,3}^2\right) e^{5 \lambda  \xi }}{16 k^4 \lambda ^4} \\
\end{array}
\right], \\
&& \tiny{ {\bf y}_3(\xi)=\left[ \begin{array}{c}
   U_3(\xi )  \\   V_3(\xi )  \\  W_3(\xi ) \\ \end{array}\right]=\left[
\begin{array}{c}
\frac{c_{2,1} \left(c_{2,1}^6+21 c_{2,2} c_{2,3} c_{2,1}^4+35 c_{2,2}^2 c_{2,3}^2 c_{2,1}^2+7 c_{2,2}^3 c_{2,3}^3\right) e^{7 \lambda  \xi }}{64
   k^6 \lambda ^6} \\
 \frac{c_{2,2} \left(7 c_{2,1}^6+35 c_{2,2} c_{2,3} c_{2,1}^4+21 c_{2,2}^2 c_{2,3}^2 c_{2,1}^2+c_{2,2}^3 c_{2,3}^3\right) e^{7 \lambda  \xi }}{64
   k^6 \lambda ^6} \\
 \frac{c_{2,3} \left(7 c_{2,1}^6+35 c_{2,2} c_{2,3} c_{2,1}^4+21 c_{2,2}^2 c_{2,3}^2 c_{2,1}^2+c_{2,2}^3 c_{2,3}^3\right) e^{7 \lambda  \xi }}{64
   k^6 \lambda ^6} \\
\end{array}
\right], } \\
&&  \cdots 
\end{eqnarray}
Likewise, symbolic calculations can be used to determine higher order correction terms. 
Thus, the series solution mentioned above can be put into the form
\begin {eqnarray}\label{ex2ex11}
&&{\bf y }(\xi)=\left[ \begin{array}{c}
   U(\xi )  \\   V(\xi )  \\  W(\xi ) \\ \end{array}\right]=\sum_{n=0}^\infty \left[ \begin{array}{c}
   U_n(\xi )  \\   V_n(\xi )  \\  W_n(\xi ) \\ \end{array}\right]= \left[ \begin{array}{c}
   \sum_{n=0}^\infty U_n(\xi )  \\   \sum_{n=0}^\infty V_n(\xi )  \\  \sum_{n=0}^\infty W_n(\xi ) \\ \end{array}\right] \nonumber \\
&& =\left[ \begin{array}{c}
   c_{2,1} \  e^{ \lambda \ \xi} +\frac{c_{2,1} \left(c_{2,1}^2+3 c_{2,2} c_{2,3}\right) e^{3 \lambda  \xi }}{4 k^2 \lambda ^2}  +   \frac{c_{2,1} \left(c_{2,1}^4+10 c_{2,2} c_{2,3} c_{2,1}^2+5 c_{2,2}^2 c_{2,3}^2\right) e^{5 \lambda  \xi }}{16 k^4 \lambda ^4}+ 
 \frac{c_{2,1} \left(c_{2,1}^6+21 c_{2,2} c_{2,3} c_{2,1}^4+35 c_{2,2}^2 c_{2,3}^2 c_{2,1}^2+7 c_{2,2}^3 c_{2,3}^3\right) e^{7 \lambda  \xi }}{64
   k^6 \lambda ^6} +  \cdots  \\

   c_{2,2} \  e^{ \lambda \ \xi} +   \frac{c_{2,2} \left(3 c_{2,1}^2+c_{2,2} c_{2,3}\right) e^{3 \lambda  \xi }}{4 k^2 \lambda ^2}+
    \frac{c_{2,2} \left(5 c_{2,1}^4+10 c_{2,2} c_{2,3} c_{2,1}^2+c_{2,2}^2 c_{2,3}^2\right) e^{5 \lambda  \xi }}{16 k^4 \lambda ^4}+ \frac{c_{2,2} \left(7 c_{2,1}^6+35 c_{2,2} c_{2,3} c_{2,1}^4+21 c_{2,2}^2 c_{2,3}^2 c_{2,1}^2+c_{2,2}^3 c_{2,3}^3\right) e^{7 \lambda  \xi }}{64
   k^6 \lambda ^6} +\cdots \\ 
     c_{2,3} \  e^{ \lambda \ \xi} + \frac{c_{2,3} \left(3 c_{2,1}^2+c_{2,2} c_{2,3}\right) e^{3 \lambda  \xi }}{4 k^2 \lambda ^2}+\frac{c_{2,3} \left(5 c_{2,1}^4+10 c_{2,2} c_{2,3} c_{2,1}^2+c_{2,2}^2 c_{2,3}^2\right) e^{5 \lambda  \xi }}{16 k^4 \lambda ^4}+
  \frac{c_{2,3} \left(7 c_{2,1}^6+35 c_{2,2} c_{2,3} c_{2,1}^4+21 c_{2,2}^2 c_{2,3}^2 c_{2,1}^2+c_{2,2}^3 c_{2,3}^3\right) e^{7 \lambda  \xi }}{64
   k^6 \lambda ^6}      +\cdots \\ \end{array}\right].  \nonumber \\
\end{eqnarray}
The closed form of the aforementioned generates series
 $$ \left(U(\xi )=\sum_{n=0}^\infty U_n(\xi ),\  V(\xi )=\sum_{n=0}^\infty V_n(\xi ),\ W(\xi )=\sum_{n=0}^\infty W_n(\xi ) \right)$$
can be produced for general values of the involve parameters since the consider equation is integrable. We employ a method utilizing multiplicative inverse of power series in the following section to obtain the closed form of the derived series for general values of relevant parameters. 
\\ 
\textbf{ $\blacksquare$  Multiplicative inverse of power series (\ref{ex2ex11}) and exact solution:  } \\ 
The series  $ \left( U(\xi )=\sum_{n=0}^\infty U_n(\xi ),\right. $ $ \left. V(\xi )=\sum_{n=0}^\infty V_n(\xi ), \right. $ $ \left. \ W(\xi )=\sum_{n=0}^\infty W_n(\xi ) \right)$ can be expressed in terms of $\zeta$ in the form $\left( U(\zeta )=  \sum_{n=0}^\infty U_n(\zeta),\right. $ $ \left. V(\zeta )=\sum_{n=0}^\infty V_n(\zeta),\right. $ $ \left. \  W(\zeta)=\sum_{n=0}^\infty W_n(\zeta) \right)$ using the substitution $ e^{ \lambda \ \xi} = \zeta $. 
Because the multiplicative inverse of a power series requires the first constant term to be non-zero, we make a small alteration and compute the multiplicative inverse of the series $\left(\frac{U(\zeta )}{ \zeta } \ ,\right. $ $ \left. \frac{V(\zeta )}{ \zeta } ,\ \frac{W(\zeta )}{ \zeta } \right) $ instead of the original power series.     
We now derive the corresponding multiplicative inverse power series as follows using the recursive formula (\ref{mi1}):
\begin {eqnarray}\label{mip0}
&&\left( \frac{U(\zeta )}{ \zeta }\right)^{-1}=  \sum_{n=0}^\infty \tilde{U}_n(\zeta) = \frac{1}{c_{2,1}}-\frac{c_{2,1}^2+3 c_{2,2} c_{2,3}}{4 k^2 \lambda ^2 c_{2,1}} \zeta ^2-\frac{c_{2,2} c_{2,3} \left(c_{2,1}^2-c_{2,2} c_{2,3}\right)}{4 k^4 \lambda ^4 c_{2,1}} \zeta ^4-\frac{c_{2,2} c_{2,3} \left(c_{2,1}^2-c_{2,2} c_{2,3}\right)^2}{16 k^6 \lambda ^6 c_{2,1}}\zeta ^6+\cdots \nonumber \\
&& \left( \frac{V(\zeta )}{ \zeta }\right)^{-1} =  \sum_{n=0}^\infty \tilde{V}_n(\zeta)= \frac{1}{c_{2,2}}-\frac{3 c_{2,1}^2+c_{2,2} c_{2,3}}{4 k^2 \lambda ^2 c_{2,2}} \zeta ^2 -\frac{c_{2,1}^2 \left(c_{2,2} c_{2,3}-c_{2,1}^2\right)}{4 k^4 \lambda ^4 c_{2,2}} \zeta ^4-\frac{c_{2,1}^2 \left(c_{2,2} c_{2,3}-c_{2,1}^2\right){}^2}{16 k^6 \lambda ^6 c_{2,2}} \zeta ^6+\cdots  \\
&& \left( \frac{W(\zeta )}{ \zeta }\right)^{-1}=  \sum_{n=0}^\infty \tilde{W}_n(\zeta) = \frac{1}{c_{2,3}}-\frac{3 c_{2,1}^2+c_{2,2} c_{2,3}}{4 k^2 \lambda ^2 c_{2,3}} \zeta ^2-\frac{c_{2,1}^2 \left(c_{2,2} c_{2,3}-c_{2,1}^2\right)}{4 k^4 \lambda ^4 c_{2,3}} \zeta ^4-\frac{c_{2,1}^2 \left(c_{2,2} c_{2,3}-c_{2,1}^2\right)^2}{16 k^6 \lambda ^6 c_{2,3}} \zeta ^6+\cdots \nonumber 
\end{eqnarray} 
Likewise, we can use symbolic computations to determine the higher order recurrence terms. The "Summing infinite series" method described in section \ref{mi} is used to determine the closed form or generating function of the aforementioned series. Either directly or by comparison with a convergent series, we determine the generating function of these infinite series. One may verify that the terms of the three series given above after $\zeta^4$ have the same ratio, i.e., 
$$\frac{\tilde{U}_{n+1}(\zeta)}{\tilde{U}_n(\zeta)}=\frac{\zeta^2 \left(c_{2,1}^2-c_{2,2} c_{2,3}\right)}{4 k^2 \lambda ^2}, \ \frac{\tilde{V}_{n+1}(\zeta)}{\tilde{V}_n(\zeta)}=-\frac{\zeta^2 \left(c_{2,1}^2-c_{2,2} c_{2,3}\right)}{4 k^2 \lambda ^2},\ \ \ \frac{\tilde{W}_{n+1}(\zeta)}{\tilde{W}_n(\zeta)}=-\frac{\zeta^2 \left(c_{2,1}^2-c_{2,2} c_{2,3}\right)}{4 k^2 \lambda ^2} \ \ n\geq 4.  $$ The above conditions lead to a geometric series with a common ratio, which is a convergent series. So the closed-form solutions of the above multiplicative inverse series are given below.
\begin {eqnarray}\label{mip6a}
\left( \frac{U(\zeta )}{ \zeta }\right)^{-1}=  \sum_{n=0}^\infty \tilde{U}_n(\zeta)&& = \frac{1}{c_{2,1}}-\frac{c_{2,1}^2+3 c_{2,2} c_{2,3}}{4 k^2 \lambda ^2 c_{2,1}} \zeta ^2-\frac{c_{2,2} c_{2,3} \left(c_{2,1}^2-c_{2,2} c_{2,3}\right)}{4 k^4 \lambda ^4 c_{2,1}} \zeta ^4 \mathbin{\slash}  \left( 1- \frac{\zeta^2 \left(c_{2,1}^2-c_{2,2} c_{2,3}\right)}{4 k^2 \lambda ^2}\right) \nonumber \\
&& =\frac{-2 c_{2,1}^2 e^{2 \lambda  \xi } \left(c_{2,2} c_{2,3} e^{2 \lambda  \xi }+4 k^2 \lambda ^2\right)+\left(c_{2,2} c_{2,3} e^{2 \lambda  \xi
   }-4 k^2 \lambda ^2\right){}^2+c_{2,1}^4 e^{4 \lambda  \xi }}{4 k^2 \lambda ^2 c_{2,1} \left(-c_{2,1}^2 e^{2 \lambda  \xi }+c_{2,2} c_{2,3}
   e^{2 \lambda  \xi }+4 k^2 \lambda ^2\right)}.\nonumber \\
 \left( \frac{V(\zeta )}{ \zeta }\right)^{-1} =  \sum_{n=0}^\infty \tilde{V}_n(\zeta)&&= \frac{1}{c_{2,2}}-\frac{3 c_{2,1}^2+c_{2,2} c_{2,3}}{4 k^2 \lambda ^2 c_{2,2}} \zeta ^2 -\frac{c_{2,1}^2 \left(c_{2,2} c_{2,3}-c_{2,1}^2\right)}{4 k^4 \lambda ^4 c_{2,2}} \zeta ^4\mathbin{\slash}  \left( 1+\frac{\zeta^2 \left(c_{2,1}^2-c_{2,2} c_{2,3}\right)}{4 k^2 \lambda ^2}\right) \nonumber  \\
 &&= \frac{-2 c_{2,1}^2 e^{2 \lambda  \xi } \left(c_{2,2} c_{2,3} e^{2 \lambda  \xi }+4 k^2 \lambda ^2\right)+\left(c_{2,2} c_{2,3} e^{2 \lambda  \xi
   }-4 k^2 \lambda ^2\right){}^2+c_{2,1}^4 e^{4 \lambda  \xi }}{4 k^2 \lambda ^2 c_{2,2} \left(c_{2,1}^2 e^{2 \lambda  \xi }-c_{2,2} c_{2,3}
   e^{2 \lambda  \xi }+4 k^2 \lambda ^2\right)}. \\ 
 \left( \frac{W(\zeta )}{ \zeta }\right)^{-1}=  \sum_{n=0}^\infty \tilde{W}_n(\zeta)&& = \frac{1}{c_{2,3}}-\frac{3 c_{2,1}^2+c_{2,2} c_{2,3}}{4 k^2 \lambda ^2 c_{2,3}} \zeta ^2-\frac{c_{2,1}^2 \left(c_{2,2} c_{2,3}-c_{2,1}^2\right)}{4 k^4 \lambda ^4 c_{2,3}} \zeta ^4\mathbin{\slash}  \left( 1+\frac{\zeta^2 \left(c_{2,1}^2-c_{2,2} c_{2,3}\right)}{4 k^2 \lambda ^2} \right) \nonumber \\
&&= \frac{-2 c_{2,1}^2 e^{2 \lambda  \xi } \left(c_{2,2} c_{2,3} e^{2 \lambda  \xi }+4 k^2 \lambda ^2\right)+\left(c_{2,2} c_{2,3} e^{2 \lambda  \xi
   }-4 k^2 \lambda ^2\right){}^2+c_{2,1}^4 e^{4 \lambda  \xi }}{4 k^2 \lambda ^2 c_{2,3} \left(c_{2,1}^2 e^{2 \lambda  \xi }-c_{2,2} c_{2,3}
   e^{2 \lambda  \xi }+4 k^2 \lambda ^2\right)}. \nonumber 
\end{eqnarray} 
Next, applying the above closed forms, the sum of the series (\ref{ex2ex11}) or $ \left(U(\xi ),\right. $ $ \left. V(\xi ),\ W(\xi ) \right)$ are given below.\\
\textbf{Solution: \ } Now using the above closed forms, we get the sum of the considered series (\ref{ex2ex11}) in the form:
\begin {eqnarray}\label{mip7}
&&{\bf y }(\zeta)=\left[ \begin{array}{c}
   U(\zeta )  \\   V(\zeta)  \\  W(\zeta) \\ \end{array}\right]=\left[ \begin{array}{c}
  \frac{\zeta}{ \left( \frac{U(\zeta )}{ \zeta }\right)^{-1}}  \\    \frac{\zeta}{ \left( \frac{V(\zeta )}{ \zeta }\right)^{-1}}   \\  \frac{\zeta}{ \left( \frac{W(\zeta )}{ \zeta }\right)^{-1}}  \\ \end{array}\right]=\left[ \begin{array}{c}
 \frac{4 k^2 \lambda ^2 c_{2,1} e^{\lambda  \xi } \left(-c_{2,1}^2 e^{2 \lambda  \xi }+c_{2,2} c_{2,3} e^{2 \lambda  \xi }+4 k^2 \lambda
   ^2\right)}{-2 c_{2,1}^2 e^{2 \lambda  \xi } \left(c_{2,2} c_{2,3} e^{2 \lambda  \xi }+4 k^2 \lambda ^2\right)+\left(c_{2,2} c_{2,3} e^{2
   \lambda  \xi }-4 k^2 \lambda ^2\right){}^2+c_{2,1}^4 e^{4 \lambda  \xi }}

  \\   
\frac{4 k^2 \lambda ^2 c_{2,2} e^{\lambda  \xi } \left(c_{2,1}^2 e^{2 \lambda  \xi }-c_{2,2} c_{2,3} e^{2 \lambda  \xi }+4 k^2 \lambda
   ^2\right)}{-2 c_{2,1}^2 e^{2 \lambda  \xi } \left(c_{2,2} c_{2,3} e^{2 \lambda  \xi }+4 k^2 \lambda ^2\right)+\left(c_{2,2} c_{2,3} e^{2
   \lambda  \xi }-4 k^2 \lambda ^2\right){}^2+c_{2,1}^4 e^{4 \lambda  \xi }}
 \\ 
 \frac{4 k^2 \lambda ^2 c_{2,3} e^{\lambda  \xi } \left(c_{2,1}^2 e^{2 \lambda  \xi }-c_{2,2} c_{2,3} e^{2 \lambda  \xi }+4 k^2 \lambda
   ^2\right)}{-2 c_{2,1}^2 e^{2 \lambda  \xi } \left(c_{2,2} c_{2,3} e^{2 \lambda  \xi }+4 k^2 \lambda ^2\right)+\left(c_{2,2} c_{2,3} e^{2
   \lambda  \xi }-4 k^2 \lambda ^2\right){}^2+c_{2,1}^4 e^{4 \lambda  \xi }}
 \\ \end{array}\right].
\end{eqnarray}
Returning to the original variable $x,t$ yields the close form solution of the series under consideration (\ref{ex2eq6}) in the form
\begin {eqnarray}\label{mip8}
&&{\bf y }(x,t)=\left[ \begin{array}{c}
   p(x,t )  \\   q(x,t ) \\  r(x,t ) \\ \end{array}\right]=\left[ \begin{array}{c}
 \frac{4 k^2 \lambda ^2 c_{2,1} e^{k \lambda  (x-3 c t)} \left(-c_{2,1}^2 e^{2 k \lambda  x}+c_{2,2} c_{2,3} e^{2 k \lambda  x}+4 k^2 \lambda ^2
   e^{2 c k \lambda  t}\right)}{-2 c_{2,1}^2 e^{2 k \lambda  (x-c t)} \left(c_{2,2} c_{2,3} e^{2 k \lambda  (x-c t)}+4 k^2 \lambda
   ^2\right)+\left(c_{2,2} c_{2,3} e^{2 k \lambda  (x-c t)}-4 k^2 \lambda ^2\right){}^2+c_{2,1}^4 e^{4 k \lambda  (x-c t)}}

  \\   
\frac{4 k^2 \lambda ^2 c_{2,2} e^{k \lambda  (x-3 c t)} \left(c_{2,1}^2 e^{2 k \lambda  x}-c_{2,2} c_{2,3} e^{2 k \lambda  x}+4 k^2 \lambda ^2
   e^{2 c k \lambda  t}\right)}{-2 c_{2,1}^2 e^{2 k \lambda  (x-c t)} \left(c_{2,2} c_{2,3} e^{2 k \lambda  (x-c t)}+4 k^2 \lambda
   ^2\right)+\left(c_{2,2} c_{2,3} e^{2 k \lambda  (x-c t)}-4 k^2 \lambda ^2\right){}^2+c_{2,1}^4 e^{4 k \lambda  (x-c t)}}
 \\ 
 \frac{4 k^2 \lambda ^2 c_{2,3} e^{k \lambda  (x-3 c t)} \left(c_{2,1}^2 e^{2 k \lambda  x}-c_{2,2} c_{2,3} e^{2 k \lambda  x}+4 k^2 \lambda ^2
   e^{2 c k \lambda  t}\right)}{-2 c_{2,1}^2 e^{2 k \lambda  (x-c t)} \left(c_{2,2} c_{2,3} e^{2 k \lambda  (x-c t)}+4 k^2 \lambda
   ^2\right)+\left(c_{2,2} c_{2,3} e^{2 k \lambda  (x-c t)}-4 k^2 \lambda ^2\right){}^2+c_{2,1}^4 e^{4 k \lambda  (x-c t)}}
 \\ \end{array}\right].
\end{eqnarray}
Using the Miura transformation (\ref{ex2ex1a}), the solution to (\ref{ex2ex1}) is now given in the form
\begin {eqnarray}\label{mip9}
&&{\bf y }(x,t)=\left[ \begin{array}{c}
   u(x,t )  \\   v(x,t ) \\  w(x,t ) \\ \end{array}\right]=\left[ \begin{array}{c}
 \frac{4 k^3 \lambda ^3 e^{k \lambda  (c t+x)} \left(c_{2,1} \left(4 k^2 \lambda ^2 e^{2 c k \lambda  t}-c_{2,2} c_{2,3} e^{2 k \lambda 
   x}\right)-4 k \lambda  c_{2,1}^2 e^{k \lambda  (c t+x)}+4 k \lambda  c_{2,2} c_{2,3} e^{k \lambda  (c t+x)}+c_{2,1}^3 e^{2 k \lambda 
   x}\right)}{\left(-4 k \lambda  c_{2,1} e^{k \lambda  (c t+x)}+c_{2,1}^2 e^{2 k \lambda  x}-c_{2,2} c_{2,3} e^{2 k \lambda  x}+4 k^2 \lambda
   ^2 e^{2 c k \lambda  t}\right){}^2}
  \\   
\frac{4 k^3 \lambda ^3 c_{2,2} e^{k \lambda  (c t+x)} \left(-c_{2,1}^2 e^{2 k \lambda  x}+c_{2,2} c_{2,3} e^{2 k \lambda  x}+4 k^2 \lambda ^2
   e^{2 c k \lambda  t}\right)}{\left(-4 k \lambda  c_{2,1} e^{k \lambda  (c t+x)}+c_{2,1}^2 e^{2 k \lambda  x}-c_{2,2} c_{2,3} e^{2 k \lambda 
   x}+4 k^2 \lambda ^2 e^{2 c k \lambda  t}\right){}^2}
 \\ 
 \frac{4 k^3 \lambda ^3 c_{2,3} e^{k \lambda  (c t+x)} \left(-c_{2,1}^2 e^{2 k \lambda  x}+c_{2,2} c_{2,3} e^{2 k \lambda  x}+4 k^2 \lambda ^2
   e^{2 c k \lambda  t}\right)}{\left(-4 k \lambda  c_{2,1} e^{k \lambda  (c t+x)}+c_{2,1}^2 e^{2 k \lambda  x}-c_{2,2} c_{2,3} e^{2 k \lambda 
   x}+4 k^2 \lambda ^2 e^{2 c k \lambda  t}\right){}^2}
 \\ \end{array}\right].
\end{eqnarray}
\textbf{Case-2.\  For Boundary Condition $U(\infty)= 0,\ V(\infty)= 0,\ W(\infty)= 0.$ }\\
Applying the boundary condition $U(\xi)\rightarrow 0,\ V(\xi)\rightarrow 0,\ W(\xi)\rightarrow 0$ as $\xi \rightarrow \infty $ for localized solution on recursive scheme (\ref{ex2ex9}) now yields $c_{2,i}=c_{3,i}=0,\ i=1,2,3$. Using these values and the processes of Case-1, we can obtain a solution that is comparable to (\ref{mip9}), with the exception of the independent variable's  change in sign. We have omitted these computations for the sake of the paper's simplicity. 
\section{Comparision of MMVP with  other methods}\label{com}
MMVP addresses linearization, recursive methods, deriving power series solutions, and obtaining exact solutions from those power series. Therefore, its methodology cannot be directly compared to the direct methods that exist for finding the exact solutions of nonlinear differential equations. However, the ultimate results of these techniques can be contrasted with the ultimate result of MMVP since both yield exact solutions. It is noteworthy that direct approaches such as the tanh method, exponential function methods, etc., rely on assuming the solutions' form, while MMVP solely depends on the boundary conditions. It can also be demonstrated that the soliton solutions derived from these techniques are specific cases of solutions produced by MMVP. Additionally, semi-analytical techniques such as ADM, HAM, and VAM provide approximate series solutions, while MMVP offers an exact solution, making direct comparisons between them impossible. However, the solutions obtained through MMVP can be utilized to verify the precision of these specified methods along with numerical approaches. The only method that can be compared in this context is RCAM.

Despite certain methodological similarities between MMVP and RCAM, the subsequent points highlight the benefits of MMVP over RCAM:
1. RCAM depends on deriving the inverse operator of the linear part; if this inverse operator cannot be obtained, RCAM fails. In contrast, MMVP does not rely on the inverse operator. However, MMVP requires knowledge of the solution to the linear part, and if this is unknown, the method cannot be applied.

2. Finding the inverse operator in RCAM is often a highly complex, and sometimes impossible, task. Even when derivable, it typically involves multiple integrations, making the process time-consuming especially for higher-order nonlinear ODEs. MMVP avoids these difficulties and does not suffer from such limitations.

3. In RCAM, computing each iteration or correction term necessitates evaluating a number of integrals equal to the order of the differential equation being solved. MMVP significantly reduces this burden by requiring only one integral evaluation per iteration or correction term.
 
\section{Advantages and Limitations of MMVP}\label{AdLim}
The MMVP offers several benefits for obtaining exact solutions:

1. It is capable of solving systems of nonlinear ordinary differential equations (SNLODEs) of any order that include multiple linear and nonlinear terms.  
2. For integrable nonlinear systems, MMVP directly yields closed-form solutions from the series expansions, whereas for non-integrable systems, closed-form solutions can be obtained by applying the multiplicative inverse of power series technique with certain parameter restrictions.  
3. The method can produce exact solutions expressed in terms of various functions such as hyperbolic, trigonometric, algebraic, and Jacobi elliptic functions.  
4. Depending on factors like the order of the system, modifications in linear terms, and initial or boundary conditions, MMVP can generate a diverse range of specific exact solutions.

However, the method also has some limitations:

1. MMVP requires knowledge of the complementary functions of the associated linear system; if these are unknown, the method cannot be applied to solve the SNLODEs.  
2. In some cases, non-trivial closed-form solutions may not be attainable from the series expansions provided by the method.

This balanced evaluation highlights MMVP’s strengths in efficiently solving many nonlinear differential systems, while also acknowledging constraints related to the availability of linear complementary functions and the form of the series solutions.
\section{Conclusion}\label{con}
An innovative approach is introduced for deriving exact solutions to nonlinear systems of ordinary differential equations. This method consists of four components. In the initial section, the examined nonlinear differential equations are simplified to a linear system of differential equations using an infinite series summation transformation and Adomian polynomials. In the second part, we introduced a revised approach to the variation of parameters utilizing an adjusted Cramer's rule for solving a linear system of block matrices. In the following section, we merged the outcomes of the prior two areas and introduced a novel recursive method for deriving a series solution of the system established in the initial section. In the final section, we discussed the fundamental ideas of the multiplicative inverse of power series and methods for summing infinite series. The effectiveness of this method is shown by successfully deriving  exact localized solutions for two coupled mKdV equation systems.

In the first example, we apply the method to a non-integrable systems of mKdV equations, typically derived from multi-layer fluid models via multiple scale expansions and the reductive perturbation method. Since the equation is non-integrable, the solutions obtained cannot be simply added to form closed-form expressions. Therefore, we employ the multiplicative inverse technique to truncate the series after a few terms under specific parameter constraints, resulting in three parameter restrictions and consequently three distinct sets of solutions.

The second example considers an integrable coupled KdV system, which has applications in shallow stratified fluids, atmospheric dynamics, and related areas. Although the MMVP method yields a complex series solution in this case, the closed-form of the series is not readily apparent. To simplify this, we apply the multiplicative inverse technique, which reveals that the derived series reduces to a geometric series after a few terms, characterized by a common ratio. This allows for summing the series and obtaining the exact closed-form solution of the integrable coupled mKdV equation.

Our results demonstrate that the proposed method efficiently produces closed-form solutions expressed via exponential functions for both integrable and non-integrable systems.  Unlike previous studies where MVP provided only approximate solutions, our enhanced MMVP approach incorporating Adomian polynomials achieves exact closed-form solutions for SNLODEs for the first time. These findings represent novel contributions not reported in existing literature. Also, a comparison of the proposed method with other methods was presented, including its advantages and limitations. In the future, we want to modify the method for partial differential equations. 
\section*{Acknowledgement}
The author is grateful and would like to sincerely thank Dr. Swaraj Paul (Assistant Professor, Allahabad University) and Dr. Soumitra Sarkar (Assistant Professor, Trivendevi Bhalotia College, Raniganj)  for their invaluable support in providing significant referred articles that greatly contributed to the development of this study.\\

\textbf{Conflict of interest:}
The author declares that he has no conflict of interest.\\

\textbf{Data availability:} Data sharing not applicable - no new data generated.\\

\textbf{Funding:}  No funds, grants, or other support was received.
\bibliographystyle{plain}
\bibliography{RefcMVP}

\begin{thebibliography}{10}

\bibitem{adomian2013solving}
George Adomian.
\newblock {\em Solving frontier problems of physics: the decomposition method},
  volume~60.
\newblock Springer Science \& Business Media, 2013.

\bibitem{brunetti2014old}
Maurizio Brunetti and A~Renato.
\newblock Old and new proofs of cramer’s rule.
\newblock {\em Applied Mathematical Sciences}, 8(133):6689--6697, 2014.

\bibitem{cao2010prolongation}
Yuan-Hao Cao and Deng-Shan Wang.
\newblock Prolongation structures of a generalized coupled korteweg-de vries
  equation and miura transformation.
\newblock {\em Communications in Nonlinear Science and Numerical Simulation},
  15(9):2344--2349, 2010.

\bibitem{das2025new}
Prakash~Kumar Das.
\newblock A new method to find exact solution of nonlinear ordinary
  differential equations: Application to derive thermophoretic waves in
  graphene sheets.
\newblock {\em arXiv preprint arXiv:2507.08808}, 2025.

\bibitem{das2018solutions}
Prakash~Kumar Das, Debabrata Singh, and MM~Panja.
\newblock {Solutions and conserved quantities of Biswas--Milovic equation by
  using the rapidly convergent approximation method}.
\newblock {\em Optik}, 174:433--446, 2018.

\bibitem{gear1984weak}
John~Anthony Gear and Roger Grimshaw.
\newblock Weak and strong interactions between internal solitary waves.
\newblock {\em Studies in Applied Mathematics}, 70(3):235--258, 1984.

\bibitem{gungor2020application}
Osman G{\"u}ng{\"o}r and Cihat Arslant{\"u}rk.
\newblock Application of variation of the parameters method for micropolar flow
  in a porous channel.
\newblock {\em Journal of Applied Mathematics and Computational Mechanics},
  19(1):17--29, 2020.

\bibitem{huang2025new}
Shaowu Huang.
\newblock A new insight into cramer’s rule via the determinants of block
  matrices.
\newblock {\em The College Mathematics Journal}, 56(1):63--64, 2025.

\bibitem{mahmood2014tan}
Anwar Ja'afar Mohamad~Jawad Mahmood, Jawad~Abu Al-Shaeer, and Abdulkarim~Rajab
  Taher.
\newblock Tan-cot function method to solve newcoupled zk and mkdv systems.
\newblock In {\em Proceedings of the World Congress on Engineering}, volume~2,
  2014.

\bibitem{mohyud2009variation}
Syed~Tauseef Mohyud-Din, Muhammad~Aslam Noor, and Asif Waheed.
\newblock Variation of parameters method for solving sixth-order boundary value
  problems.
\newblock {\em Communications of the Korean Mathematical Society},
  24(4):605--615, 2009.

\bibitem{moore2014application}
Travis~J Moore.
\newblock {\em Application of variation of parameters to solve nonlinear
  multimode heat transfer problems}.
\newblock Brigham Young University, 2014.

\bibitem{moore2019comparison}
Travis~J Moore and Vedat~S Ert{\"u}rk.
\newblock Comparison of the method of variation of parameters to
  semi-analytical methods for solving nonlinear boundary value problems in
  engineering.
\newblock {\em Nonlinear Engineering}, 9(1):1--13, 2019.

\bibitem{noor2008variation}
A~Waheed.
\newblock Variation of parameters method for solving fifth-order boundary value
  problems.
\newblock {\em Appl. Math. Inf. Sci}, 2(2):135--141, 2008.

\bibitem{wazwaz2012study}
Abdul-Majid Wazwaz.
\newblock A study on two coupled modified kdv systems with time-dependent and
  constant coefficients.
\newblock {\em Palestine Journal of Mathematics}, 1:38--48, 2012.

\bibitem{Zemlyanukhin}
Aleksandr Zemlyanukhin, Nikolay Artamonov, Andrej Bochkarev, and Vladimir
  Bezlyudny.
\newblock Power series reversion and exact solutions of nonlinear mathematical
  physics equations.
\newblock {\em Izvestiya VUZ. Applied Nonlinear Dynamics}, 06 2025.

\end{thebibliography}
\end{document}